\documentclass[onecolumn,preprintnumbers,nofootinbib]{revtex4-1}
\usepackage{sidecap}
\usepackage{wrapfig}

\usepackage{graphicx}  
\usepackage{amssymb}
\usepackage{color}
\usepackage{enumerate}
\usepackage{amsmath}
\usepackage{slashed}
\usepackage{comment}

\def\beq{\begin{equation}}
\def\eeq{\end{equation}}
\def\beqn{\begin{eqnarray}}
\def\eeqn{\end{eqnarray}}

\renewcommand{\texttt}{{}}
\newcommand{\be}{\begin{eqnarray}}
\newcommand{\ee}{\end{eqnarray}}

\begin{document}

\title{Spacetime completeness of non-singular black holes in conformal gravity}

\author{Cosimo Bambi}
\email{bambi@fudan.edu.cn}
\affiliation{Center for Field Theory and Particle Physics and Department of Physics, 
Fudan University, 200433 Shanghai, China}
\affiliation{Theoretical Astrophysics, Eberhard-Karls Universit\"at T\"ubingen, 72076 T\"ubingen, Germany}

\author{Leonardo Modesto}
\email{lmodesto@sustc.edu.cn, lmodesto1905@icloud.com}
\affiliation{Department of Physics, Southern University of Science and Technology, Shenzhen 518055, China} 
\affiliation{Center for Field Theory and Particle Physics and Department of Physics, 
Fudan University, 200433 Shanghai, China}

\author{Les\l{}aw Rachwa\l{}}
\email{grzerach@gmail.com}
\affiliation{Center for Field Theory and Particle Physics and Department of Physics, 
Fudan University, 200433 Shanghai, China}

\begin{abstract} \noindent
We explicitly prove that the Weyl conformal symmetry solves the black hole singularity problem, otherwise unavoidable in a generally covariant local or non-local gravitational theory. Moreover, we yield explicit examples of local and non-local theories enjoying Weyl and diffeomorphism symmetry (in short co-covariant theories). Following the seminal paper by Narlikar and Kembhavi, we provide an explicit construction of singularity-free spherically symmetric and axi-symmetric exact solutions for black hole spacetimes conformally equivalent to the Schwarzschild or the Kerr spacetime. We first check the absence of divergences in the Kretschmann invariant for the rescaled metrics. Afterwords, we show that the new types of black holes are geodesically complete and linked by a Newman-Janis transformation just as in standard general relativity (based on Einstein-Hilbert action). Furthermore, we argue that no massive or massless particles can reach the former Schwarzschild singularity or touch the former Kerr ring singularity in a finite amount of their proper time or of their affine parameter. Finally, we discuss the Raychaudhuri equation in a co-covariant theory and we show that the expansion parameter for congruences of both types of geodesics (for massless and massive particles) never reaches minus infinity. Actually, the null geodesics become parallel at the $r=0$ point in the Schwarzschild spacetime (the origin) and the focusing of geodesics is avoided. The arguments of regularity of curvature invariants, geodesic completeness, and finiteness of geodesics' expansion parameter ensure us that we are dealing with singularity-free and geodesically-complete black hole spacetimes. 
\end{abstract}

\maketitle

\tableofcontents


\section{Introduction}

Einstein's theory of gravity is an excellent classical two-derivative theory based on general covariance. However, almost all  solutions of Einstein's equations of motion (EOM),\footnote{In this paper,  we will employ natural units in which $c = \hbar = k_{\rm B} = 1$ and adopt a metric signature $(-+++)$,
but in all plots we will use Planck units $M_P= \sqrt{\hbar c/G_N} = 1$. By bold characters we denote tensors without writing all their indices.  
Moreover, we put a hat over quantities computed using the hatted metric $\hat g$.
Finally, by $M$ we actually mean the geometrical mass $G_N  M$.} 
\be
{\bf G} = 8 \pi G_N \, {\bf T}, 
\label{EE}
\ee
manifest spacetime singularities. Many attempts have been done to overcome this issue, but no one has been completely successful. In particular, in higher derivative local~\cite{Stelle, shapiro3, Shapirobook, HigherDG} or non-local~\cite{Efimov} gravitational theories, the Newtonian potential for point-like sources turns out to have a universal regular and constant behaviour near $r=0$~\cite{Tiberio, Giacchini:2016xns}. 
Other regular spacetimes describing the gravitational collapse and black holes  
has been derived and studied in \cite{frolov, Frolov+-1,Frolov+,Frolov+2,Frolov+3,Frolov+4, Frolov:2016pav, Frolov:2016xhq} and \cite{ModestoMoffatNico, BambiMalaModesto2, BambiMalaModesto, 
DeLorenzo:2014pta, Bambi:2016uda,calcagnimodesto}.
However, most of the singularities of Einstein's gravity are still present in the weakly non-local or local higher derivative gravitational theories~\cite{exactsol}, and we believe that only a new symmetry principle together with general covariance can finally rid of the spacetime singularities. While in a non-local theory we may have singularity-free cosmological solutions~\cite{koshe1}, in general the singularities cannot be wiped away by engineering the action; it must be a special symmetry principle of Nature to make them harmless. We think that the scale-invariant structure  at short distances~\cite{Narlikar, Englert, thooft0, thooft, Penrose, Mannheim, barsTurok, Bars2, Prester:2013fia, CGLL} could be crucial in this respect.

In this paper, we do not focus on a particular conformally invariant gravitational theory, but we address the issue of singularities in a general theory invariant under Weyl conformal transformations. Therefore,  we show that the resolution of spacetime singularities is more related to the symmetry of a gravitational theory than on its precise dynamical structure. In a $D$-dimensional spacetime, the most popular two-derivative conformally invariant theory is Einstein's conformal gravity. The theory is constructed replacing the metric $g_{\mu\nu}$ with the auxiliary dilaton field $\phi$ and the metric $\hat{g}_{\mu\nu}$, defined as~\cite{Englert, thooft0}
\be
g_{\mu\nu} = \left( \phi^2 \kappa_D^2 \right)^{\frac{2}{D-2}} \hat{g}_{\mu\nu} \, ,
\label{phighat}
\ee
in the Einstein-Hilbert gravitational action $- 2 \kappa_D^2\int\!d^Dx\sqrt{|g|} R(g)\,$. 
Here $2/\kappa_D^2=\frac1{16 \pi G_N}$. 
The Lagrangian density of the outcome reads
\be
\mathcal{L}_{\rm g} 
= -  
2  \, \sqrt{ |\hat{g}| } 
\left[   \phi^2 R(\hat{g}) +
\frac{4(D-1)}{ D-2} \hat{g}^{\mu\nu} (\partial_\mu \phi) (\partial_\nu \phi) \right] .
\label{CEHG}
\ee
There exist two other examples of local four-derivative conformally invariant gravitational theories in $D=4$, namely their Lagrangians are \cite{Mannheim, conformalComp, Fradkin:1981iu}
\be
&& \mathcal{L}_1 = a\,{\bf C}^2 + b\,{\bf Riem}^*\,{\bf Riem} \, , \label{CFM}\\
&&  \mathcal{L}_2  = 12 \phi \left( \Box - \frac{1}{6} R \right) \phi + \lambda \phi^4 + a\,{\bf C}^2 + b\,{\bf Riem}^*\,{\bf Riem} \, ,
\label{C2FT}
\ee
where ${\bf C}$ is the Weyl (conformal) curvature tensor and ${\bf Riem}$ is the standard Riemann curvature tensor\footnote{We notice that the term with dual Riemann tensor ${\bf Riem}^*$ contracted with Riemann tensor is actually a total derivative and vanishes upon integration over non-compact manifolds without boundaries, hence it is trivially conformally symmetric.}. The Lagrangians~(\ref{CEHG}), (\ref{CFM}), and (\ref{C2FT}), for any values of the real parameters 
$a, b$, and $\lambda$, are useful examples for understanding the connection between conformal symmetry and singularities. However, the theory in~(\ref{CEHG}) is non-renormalizable at quantum level in $D>2$, while those in~(\ref{C2FT}) are renormalizable but not finite (contain UV divergences already at one loop), implying that the Weyl symmetry is anomalous and cannot be preserved at quantum level. Moreover, the theories in the last two sets are non-unitary because of the presence of ``bad" ghosts in the spectrum. We invite the reader to look at \cite{solveghost} for some attempts to overcome this difficulty.

Recently, it has been shown that a class of weakly non-local gravitational theories is anomaly-free~\cite{CGLL}. This is a non-trivial achievement based on the UV-finiteness of the theory at quantum level without and with supersymmetry~\cite{kuzmin, modestoLeslaw, universality, KoshelevStaro, entanglement, GiaccariSugra} as an extent of the super-renormalizability property of such theories~\cite{Krasnikov, Tombo, Tomboulis:2015gfa, Briscese:2013lna, Khoury, modesto, Mtheory, GiaccariSugra, M3, M4}. In spite of the quite involved non-local structure of the theory, the main lesson is that conformal invariance can be realized in Nature~\cite{Fradkin:1981iu, PercacciConf, Percacci2, Percacci3} consistently with quantum field theory~\cite{CGLL} and unitarity~\cite{Khoury, Dona, Cnl1, Modesto:2013jea}. The recent results about conformal invariance could be likely exported to a class of local Lee-Wick \cite{LW} gravitational theories super-renormalizable or finite at quantum level~\cite{Shapiro:2015uxa,Modesto:2015ozb, Modesto:2016ofr, Accioly:2016etf,Accioly:2016qeb}. Finally, we know that at the macroscopic scales accessible to gravitational experiments conformal symmetry is not realized. On the other hand we know that in high energy physics phenomenology conformal invariance could exist only in the high energy regime, where all particles are effectively massless. Therefore, in the theory we must somehow find a mechanism to end up with a low-energy effective action without explicit Weyl symmetry. Two possible ways to achieve this purpose are the following. The Weyl symmetry is spontaneously broken, or it is realized in the UV regime at a trivial or a non-trivial (interacting) UV fixed point \cite{Reuter:1996cp, Niedermaier:2006wt, Codello:2008vh, Codello:2015oqa}. In the latter case the black holes' physics described in this paper takes place at the UV fixed point where the Weyl symmetry turns out to be an  exact symmetry.


\section{Spacetime Singularities in Conformal Gravity \label{sec-2}}

In the past, many attempts have been worked out to avoid the singularity problem in general relativity. However, conformal invariance seems to be the unique way to get rid of spacetime singularities in a gravitational theory. One can quite easily convince himself that in a conformally invariant theory there are no singularities in Friedman-Robertson-Walker (FRW) spacetimes. Indeed, all FRW spacetimes are equivalent, by a conformal transformation, to the Minkowski spacetime, which is everywhere regular, of course. The Weyl tensor is identically zero for an FRW spacetime (all FRW spacetimes are conformally flat), but the general quantities needed to be studied to infer the regularity of the spacetime are constructed out of the Riemann tensor (this will be clear at the end of this section). 

Less trivial is the case of black hole singularities and numerous attempts have been done in this 
direction~\cite{thooft,barsTurok, Bars2}. In this paper, we would complete the studies well displayed and developed by Narlikar and Kembhavi~\cite{Narlikar} to include also the Schwarzschild and the Kerr metrics in their list of singularity-free spacetimes. We now remind here the logic introduced by the two authors, but first we discuss the equations of motion (EOM) of a general conformally invariant gravitational theory and their exact vacuum solutions. To make our discussion more precise we will concentrate on the example of conformal Einstein's theory \eqref{CEHG}, where the conformal compensator (dilaton) field must be present.

First one derives the EOM $E_{\mu\nu}$ and $E_\phi$ for a favourite conformally invariant theory, which we explicitly define as follows, 
\be
E_{\mu\nu} =\frac{\delta S_{\rm conf}}{\delta \hat{g}_{\mu\nu}}=0\quad{\rm and}\quad E_\phi=\frac{\delta S_{\rm conf}}{\delta \phi}=0\,.
\ee
The solution of them is a pseudo-Riemannian spacetime manifold $\mathcal{M}$ equipped with a metric tensor $\hat{g}_{\mu\nu}$ and a scalar field $\phi$ as defined in~(\ref{phighat}). For a general theory, in which all the operators resulting from the variation of the action $S_{\rm conf}$ with respect to the metric $\hat{g}_{\mu\nu}$ are at least linear in the Ricci tensor $\bf Ric$ (notice that this is ${\bf Ric}$ and not ${\bf \hat{R}ic}$), the EOM are defined implicitly as 
\be
&& E_{\mu\nu} ={\mathcal E}_{\mu\nu}^{\alpha \beta} R_{\alpha \beta} (g)= 0 \, , \quad \mbox{where} \quad g_{\mu\nu} = \left( \phi^2 \kappa_D^2 \right)^{\frac{2}{D-2}} \hat{g}_{\mu\nu} 
\quad {\rm and} \quad \mathcal{E}_{\mu\nu}^{\alpha \beta} ={\mathcal E}_{\mu\nu}^{\alpha \beta} (R, {\bf Ric}, {\bf Riem}, \Box, \nabla)  \, , 
 \label{ConfEOMg} 
\\
&& 
 E_\phi=\hat{\Box} \phi = \frac{D-2}{4(D-1)} \hat{R} \phi 
 + \dots \,  .
 \label{ConfEOMPhi}
\ee
In \eqref{ConfEOMg} ${\mathcal E}_{\mu\nu}^{\alpha \beta} (R, {\bf Ric}, {\bf Riem}, \Box, \nabla)$ is some tensor constructed with $R$, ${\bf Ric}$, ${\bf Riem}$, and/or $\nabla$ and $\Box$ operators. The dots in~(\ref{ConfEOMPhi}) stand for higher derivative operators at least linear in 
${\bf Ric}$ or $R$, which are zero for Ricci-flat spacetimes. Therefore, in this class of theories all the vacuum solutions present in E-H gravity (like Schwarzschild or Kerr metric) are also exact solutions of the conformally invariant theory. And the following implications turn out to be correct, 
\be
\!\!\!
g_{\mu\nu} = 
(\phi \, \kappa_D)^{\frac{4}{D-2}} \, \hat{g}_{\mu\nu} \quad {\rm and} 
\quad \phi = \kappa_D^{-1}  
\,\,\, \Longrightarrow \,\,\, 
R_{\mu\nu} \! \left(  
(\phi \, \kappa_D)^{\frac{4}{D-2}} \, \hat{g}_{\mu\nu} \right) = 0 \,\,\, \Longrightarrow \,\,\, 
E_{\mu\nu} \! \left( (\phi \, \kappa_D)^{\frac{4}{D-2}} \, \hat{g}_{\mu\nu} \right) = 0 \, . 
\label{implication}
\ee
This is the case for theories \eqref{CEHG}, \eqref{CFM} and \eqref{C2FT} in $D=4$ and anomaly-free theory in \cite{CGLL}. Notice that if the contrary implication to \eqref{implication} is true for a non-local theory then the spectrum of this theory is the same as of the local Einstein-Hilbert theory at the classical non-perturbative level.

The EOM \eqref{ConfEOMg} and \eqref{ConfEOMPhi} are by construction conformally invariant, hence if we consider another manifold $\mathcal{M}^*$ obtained from $\mathcal{M}$ by a conformal transformation
\be
 \hat{g}_{\mu\nu}^{*} = \Omega^2 \, \hat{g}_{\mu\nu} \,  ,\quad 
 \phi^{*} = \Omega^{\frac{2-D}{2}} \, \phi \, ,
 \ee
then also $\hat{g}_{\mu\nu}^{*}$ and $\phi^{*}$ satisfy the EOM. The transformation $\hat g_{\mu\nu} \to \hat g^*_{\mu\nu}$ and $\phi \rightarrow \phi^*$ is mathematically valid provided $\Omega^{-1}$ does not vanish (or become infinite) \cite{Narlikar}. It is assumed that $\Omega=\Omega(x)$ is a twice differentiable function of the spacetime 
coordinates with demand that
\be
0 < \Omega < +\infty \, .
\ee
It was then shown in~\cite{Narlikar} that the manifolds $\mathcal{M}^*$ for the Belinskii, Khalatnikov \& Lifshitz (BKL) and for the Taub-Nut metrics are geodesically complete while the original manifolds $\mathcal{M}$ are not. Notice that here we have changed notation with respect to the original paper~\cite{Narlikar}, namely for us the regular manifold is $\mathcal{M}^*$.

\section{Singularities and observables in conformal gravity}

In this section we address two main problems in conformal gravity, namely: 
how we can check the regularity of a spacetime, and how the Universe gets out the Weyl conformal invariant phase.

\subsection{Regularity of the spacetime}

It is often thought that the regularity of a spacetime can be investigated by studying the regularity of operators invariant under the symmetries of the theory. However, in general, this approach is incorrect. Note that even the famous theorems by Hawking and Penrose are about geodesic singularities, rather than about curvature singularities. In this paper we prove that our black hole exact solutions are regular on the basis of the geodetic completion of the spacetime. Nevertheless, we also check the regularity of curvature operators invariant only under general coordinate transformations.

Let us start addressing the singularities' issue from the point of view of the curvature invariants. In the case of conformally invariant theories, we should find operators invariant under the symmetry group Weyl$\times$Diff. 
If the theory involves the dilaton field $\phi$, like in the theory $\mathcal{L}_{\rm g}$ in (\ref{CEHG}), it is incorrect to investigate the regularity of the spacetime with operators explicitly containing such field  because they mix together the metric and the dilaton: if an invariant involves two or more fields, we can only infer about the properties of composite operators, but nothing about the constituent fields. In the case of Weyl gravity (the Lagrangian is $\mathcal{L}_1$ in (\ref{CFM})) there is no dilaton and the curvature invariants can only be functions of the metric tensor $\hat{g}_{\mu\nu}$. Nevertheless, we cannot construct any operator that is at the same time Weyl and Diff invariant in Weyl gravity. So we cannot study any invariant operator to check if the spacetime is singularity free.

However, we can infer about the regularity of the spacetime making use of the gauge transformations at our disposal and studying the geodesic completion of the resulting spacetime. In the symmetric phase we can make a coordinate transformation or a Weyl transformation or both. If with a group element of Weyl$\times$Diff we are able to turn a singular metric in a regular one, then there is no singularity. This is exactly what happens for the coordinate singularity at the event horizon (EH) in Einstein's gravity. Indeed, we can express the Schwarzschild metric in, for example, Gullstand-Painlev\'e coordinates 
and the singularity at the EH disappears (note that the coordinate transformation is singular at the EH.) Similarly in conformal gravity we can make a Weyl rescaling and the new black hole metric turns out to be regular everywhere as can be explicitly proved looking at the geodesic completion. In other words in both cases the singularity is just an artifact of the gauge. 
We can here explicitly see how crucial the symmetries are in removing the singularities.
As already mentioned, given the metric we can infer about the regularity studying the geodesic completion of the manifold. 
In practical terms, we should analyze the propagation of massive, massless, and conformally coupled particles in the Weyl rescaled spacetime. Note that the dynamics for conformally couple particles involves also the dilaton field. 
At the same time we can also check the regularity of the rescaled metric using Diff invariant operators, namely the Ricci scalar and the Kretschmann invariant. Moreover, for the rescaled Schwarzschild metric any curvature invariant will turn out to be regular everywhere because the two-dimensional transverse area never becomes zero, but bounces back to an infinite value when approching $r=0$.

Now the physical issue is how to select out one specific metric in an infinite class of regular (and only regular) spacetime. This is exactly what we are going to address in the next subsection.

\subsection{Broken and unbroken phases of conformal symmetry} 

The world around us is not conformally invariant. If conformal symmetry is a fundamental symmetry in Nature, it must be somehow broken. The situation may be similar to the electroweak symmetry in the Standard Model of particle physics: today the electroweak symmetry is broken, but in the early Universe, when the temperature of the primordial plasma was higher than $\sim 100$~GeV, we were presumably in the symmetric phase.

In the symmetric phase, the theory is explicitly invariant under conformal transformations. This means, in particular, that all physical quantities cannot depend on the choice of the gauge, namely on the conformal factor $\Omega$. Note, however, that the physical quantities are different in the symmetric and in the broken phases. Indeed, in the symmetric phase it is not possible to perform any measurement of lengths and time intervals because it is not possible to define a standard rod and a standard clock. We do not have any notion of proper time because massive particles are not allowed and massless particles cannot serve as clocks. However, this feature of the conformal phase is not
inconsistent with the analysis of singularities developed in the previous subsection that is purely based on mathematical ground.  There are no doubts that the singularity is not physical, but the problem is how to select one particular metric in a class (probably infinity) of regular spacetimes.

In the spontaneously broken phase, 
it is Nature to select somehow a specific vacuum. Again, this is analog to the electroweak symmetry breaking in the Standard Model of particle physics. Some ideas on how the symmetry can be spontaneously broken are presented in~\cite{Englert,thooft,thooft0,Deser:1970hs,Englert:1975wj}. Now different vacua can have different physical properties because they represent different configurations: conformal symmetry is broken and, therefore, the choice of the conformal factor $\Omega$ produces observable effects (see, e.g., Ref.~\cite{zhengcao}).
However, it is crucial that Nature can only select a regular vacuum solution in a large class of vacuum solutions and not a singular one (see previous subsection). Moreover, as we will stress even later, there is no fine-tuning in this choice because the whole class of regular black hole solutions share the same good properties.


\section{Avoiding the Schwarzschild singularity in conformal gravity}

In the first part of this section we explicitly construct a class of singularity-free spherically symmetric black hole solutions, and we show the regularity of the spacetime evaluating the Ricci scalar and the Kretschmann invariant for the metric $\hat{g}_{\mu\nu}$. In the second subsection we will study the geodesic completion of the $\mathcal{M}^*$ spacetime.

\subsection{Non-singular spherically symmetric black hole} 

The Schwarzschild metric is an exact solution of the EOM~(\ref{ConfEOMg}) and (\ref{ConfEOMPhi}), and can be explicitly written in terms of $\phi$ and $\hat{g}_{\mu\nu}$, i.e. 
\be
 g_{\mu\nu}^{{\rm Schw}} = (\phi \, \kappa_D)^{\frac{4}{D-2}} \, \hat{g}_{\mu\nu} \, .
\ee
However, we can rescale both the scalar field $\phi$ and the metric $\hat{g}_{\mu\nu}$, namely 
 \be
 g_{\mu\nu}^{{\rm Schw}} = (\phi \, \kappa_D)^{\frac{4}{D-2}} \, \hat{g}_{\mu\nu} = 
 (\phi^* \, \kappa_D)^{\frac{4}{D-2}} \, \hat{g}^*_{\mu\nu}
 \quad \Longleftrightarrow \quad 
 \hat{g}_{\mu\nu}^{*} = \Omega^2 \, \hat{g}_{\mu\nu} \,  ,\quad 
 \phi^{*} = \Omega^{\frac{2-D}{2}} \, \phi . 
 \ee

For $\phi= \kappa_D^{-1}$ and $\Omega=1$, we get the Schwarzschild spacetime $\hat{g}_{\mu\nu}^* = \hat{g}_{\mu\nu}=g_{\mu\nu}^{{\rm Schw}}$. By making use of the conformal rescaling, we can construct an infinite number of exact solutions conformally equivalent to the Schwarzschild metric. Moreover, 
 \be
 R_{\mu\nu}(g_{\mu\nu}^{{\rm Schw}}) = 0 \quad \Longrightarrow \quad 
 \hat{R}_{\mu\nu}(\hat{g}^*_{\mu\nu}) \neq 0 \, ,
 \ee
therefore conformally equivalent solutions are not Ricci-flat.

We now explicitly provide -- in whatever conformally invariant theory -- an example of singularity-free exact black hole solution obtained by rescaling the Schwarzschild metric by a suitable overall warp factor $\Omega$ (see also~\cite{Prester:2013fia} for the use of similar methods). For the sake of simplicity, here we assume $D=4$. The new singularity-free black hole metric looks like (later in this section we will prove the regularity of the spacetime)
\be
&& ds^{* 2} \equiv 
\hat{g}_{\mu\nu}^* dx^\mu dx^\nu = S(r) \hat{g}_{\mu\nu} dx^\mu dx^\nu
= S(r) \left[ \left( 1- \frac{2 M}{r} \right) dt^2 + \frac{dr^2}{1- \frac{2 M}{r}} + r^2 d \Omega^{2} \right] \,  , \label{NRBH} \\
&& \phi^* = S(r)^{-1/2} \kappa_4^{-1} \, , 
\ee
where the following conformal factor $\Omega^2=S$ depending only  on the radial Schwarzschild coordinate $r$ is 
\be
S(r) = \frac{1}{r^2} \left( \frac{L^4}{r^2}+r^2 \right) \, .
\label{grazieMode}
\ee
Here $L$ is a length scale introduced for dimensional reason. It could be equal to the Planck length $L = L_P$, to a possible other fundamental scale of the theory, or even $L \propto M$. The first two options are realized with the scales already present in the theory, while the last one is with the scale that breaks conformal symmetry on-shell.

The scale factor $S(r)$ given in~(\ref{grazieMode}) meets the conditions $S^{-1}(0) =0$ and $S^{-1}(\infty) = 1$. Moreover the Schwarzschild singularity (in $r=0$) appears exactly where the conformal transformation becomes singular, i.e. where $S^{-1} = 0$. 
Here we must understand the singularity issue as an artifact of the conformal gauge. The situation is exactly the same as with the FRW spacetime. There the scale factor is singular at the time of the Big Bang (and this is classically thought of as the singularity moment), but still the conformally equivalent metric is flat and regular everywhere and for any time. This is the way how -- thanks to conformal symmetry -- the Big Bang singularity is resolved. In the case of black hole solutions, we map the spacetime $\cal M$ into a regular $\cal M^*$ via a conformal rescaling.

Of course there is an infinite class of such functions $S(r)$ that enable us to map the singular Schwarzschild spacetime in an everywhere regular one. Here we concentrate on one example with the smallest possible exponent of $r$.

The line element of the metric with the scale factor~(\ref{grazieMode}) reads 
\be
ds^{*2} =  -\frac{1}{r^2} \left( \frac{L^4}{r^2}+r^2 \right) \left( 1 - \frac{2M}{r} \right) dt^2 
+  \frac{1}{r^2} \left( \frac{L^4}{r^2}+r^2 \right) \frac{dr^2}{1 - \frac{2M}{r} } +
\left( \frac{L^4}{r^2}+r^2 \right) d \Omega^{2}  \, .  
\ee
The Kretschmann invariant ${ \hat{K} } = {\bf \hat{R}iem}^2$ is also simple and can be displayed here,
\be
&& { \hat{K}} = \frac{16 r^2}{\left(L^4+r^4\right)^6} \times \left[ 
 L^{16} \left(39 M^2-20 M r+3 r^2\right)+2 L^{12} r^4
   \left(66 M^2-32 M r+3 r^2\right) \right. \nonumber \\
 && \hspace{3cm} 
  \left. + L^8 r^8 \left(342 M^2-284 M r+63
   r^2\right)+12 L^4 M^2 r^{12}+3 M^2 r^{16}  \right]  ,
\ee
while the Ricci scalar reads 
\be
\hat{R} = -\frac{12 L^4 r \left(L^4 (r-4 M)+r^4 (3 r-8 M)\right)}{\left(L^4+r^4\right)^3} \, .
\ee
Therefore, ${ \hat{K}}$ and $\hat{R}$ are regular for all  $r$. Last, the Hawking temperature remains unchanged, namely $T_H = \frac{1}{8 \pi  M}$, because it is invariant under a conformal rescaling of the metric.

There is another scale factor that captures the same properties as of~(\ref{grazieMode}), but will simplify the analysis of the geodesic completion, namely 
\be
S(r) = \left(1 + \frac{L^2}{r^2}\right)^2 .
\label{geoS}
\ee 
Now, the Kretschmann invariant reads, 
\be
&& { \hat{K} }= \frac{1}{\left(L^2+r^2\right)^8} \times 
16 \,  r^2 \left[  L^8 \left(39 M^2-20 M r+3 r^2\right)+2 L^6 r^2 \left(42 M^2-16
   M r+r^2\right)  \right. \nonumber \\
   && \hspace{3.8cm}
    \left. +L^4 r^4 \left(150 M^2-108 M r+23 r^2\right)+12 L^2 M^2 r^6+3 M^2
   r^8 \right],
\ee
which is regular everywhere, including at $r=0$.

\begin{figure}
\begin{center}
\includegraphics[type=pdf,ext=.pdf,read=.pdf,height=7.0cm]{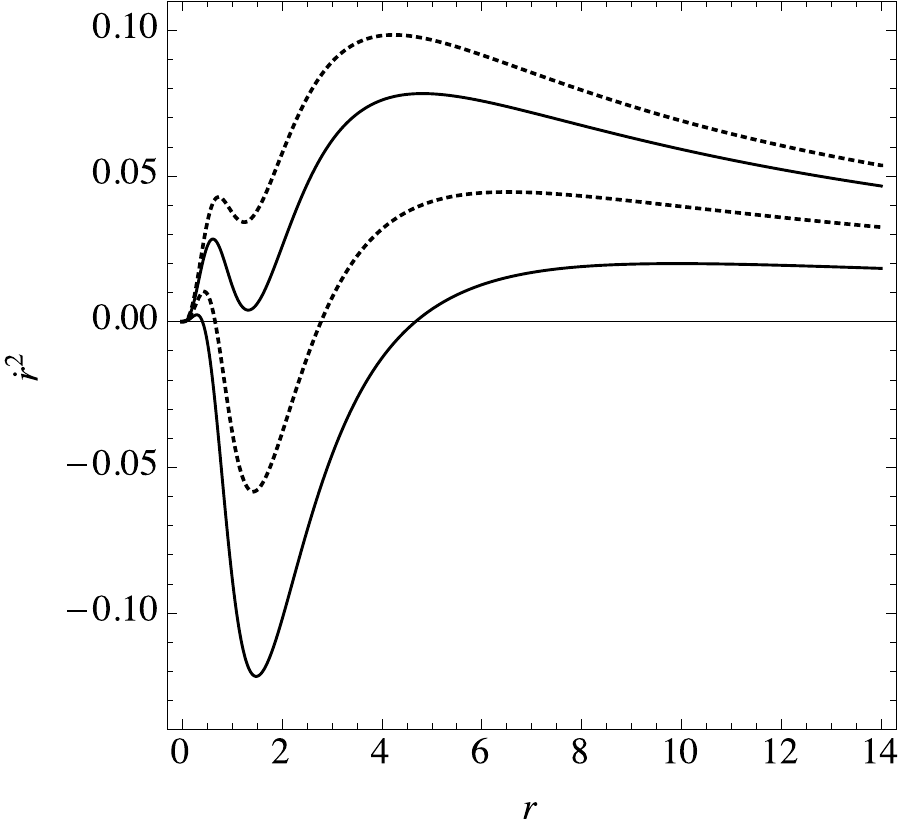}
\end{center}
\caption{The $\dot{r}^2$ quantity as a function of the radial coordinate $r$ for $M=0.2$, 0.3, 0.4, and 0.45 (from the bottom to the top), $L=1$, and $e=1$. \label{rdotMSchw}}
\end{figure}

\begin{figure}
\begin{center}
\includegraphics[type=pdf,ext=.pdf,read=.pdf,height=7.0cm]{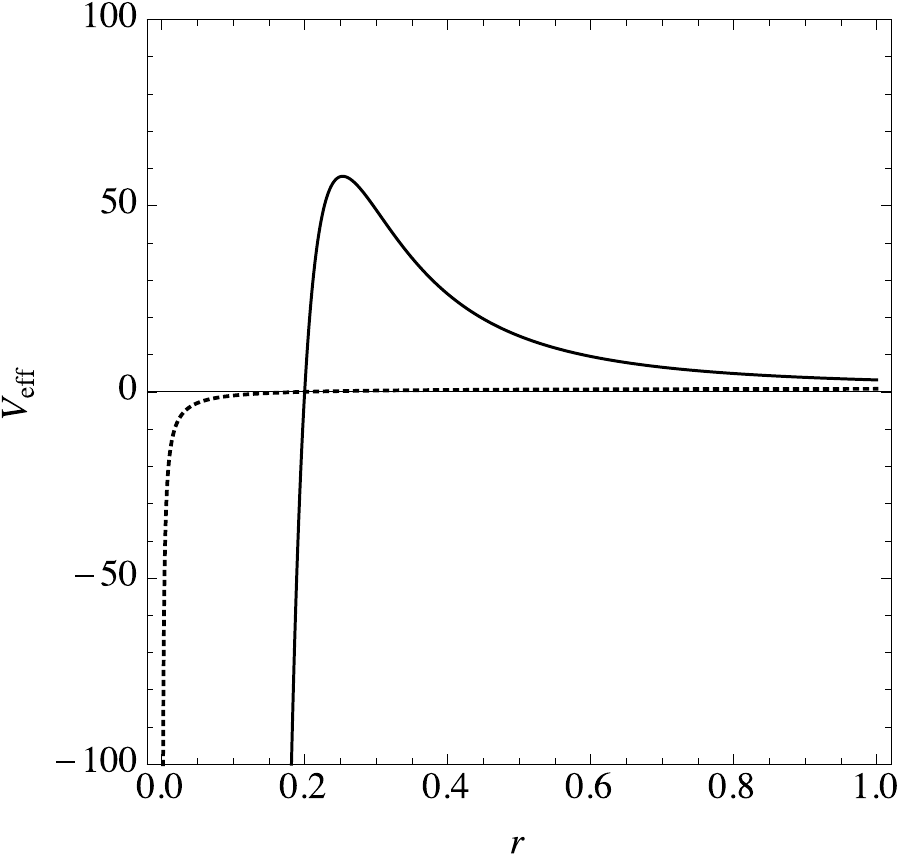}
\end{center}
\caption{Effective potential $V_{\rm eff} = - \hat{g}_{tt} = S(r) (1-2M/r)$ as a function of the radial coordinate $r$ for $M=0.1$ and $L=1$ (solid curve) compared to the Schwarzschild one (dashed curve). \label{rdotMSchw2}}
\end{figure}

\subsection{Geodesic completion} 

In this section we show the geodesic completion of the conformally rescaled Schwarzschild metric using three different probes. We will use in turn a massive particle, a conformally coupled massive particle, and a massless particle. For the sake of simplicity, we rename the regular metric by $\hat{g}_{\mu\nu}$ in place of $\hat{g}^*_{\mu\nu}$.

\subsubsection{Non-conformally coupled massive particle probe}

Let us now discuss the issue of geodesic completeness of spacetime manifolds in conformal gravity. We will focus on the geodesic motion of some probe material point in the spacetime whose metric is given by~\eqref{NRBH}. We now show that any probe  massive particle cannot fall into $r=0$ in a finite proper time. We will later study the motion of a test material point conformally coupled to conformal gravity, but the outcome will be essentially the same. We consider  radial geodesic motion for a massive test-particle
  \begin{eqnarray}
\hat{g}_{tt} \dot{t}^2 + \hat{g}_{rr} \dot{r}^2 = - 1 \, , \quad
E = - m \hat{g}_{tt} \dot{t}
   \quad \Longrightarrow \quad
 (- \hat{g}_{tt} \, \hat{g}_{rr}) \dot{r}^2 = \frac{E^2}{m^2} + \hat{g}_{tt} \, ,
  \label{geometricabella}  
  \end{eqnarray}
where the dot $\dot{}$ stands for the derivative of the proper time $\tau$, and $E$ and $m$ are, respectively, the energy and the rest-mass of the test-particle. 

For the sake of simplicity from now on we identify the regular rescaled metric with $\hat{g}_{\mu\nu}$ unlike the notation that we introduced formerly, namely $\hat{g}^*_{\mu\nu}$.

If the particle falls from spatial infinity with zero initial radial velocity, the energy is the rest-mass of the particle, $E=m$. We can write~(\ref{geometricabella}) in a more familiar form, which makes use of an effective potential $V_{\rm eff}$, namely
    \begin{eqnarray}
    \hspace{-0.7cm} 
    \overbrace{(- \hat{g}_{tt} \, \hat{g}_{rr})}^{\geqslant 0 \,\, \forall r} 
  \dot{r}^2 + \overbrace{V_{\rm eff}}^{- \hat{g}_{tt}}(r) = \frac{E^2}{m^2} 
  \quad \Longrightarrow \quad 
   S(r)^2 \dot{r}^2 +S(r) \left( 1 - \frac{2 M}{r} \right) = \frac{E^2}{m^2} \, , \quad  \dot{r}^2 +\frac{1}{S(r)} \left( 1 - \frac{2 M}{r} \right) = \frac{e^2}{S(r)^{2}} \, ,  
  \label{geometricabella2}  
  \end{eqnarray}
  where we have introduced $e = E/m$. Equation~(\ref{geometricabella2}) can be rewritten as
  \be
   \dot{r}^2  = \frac{e^2 - S(r) \left( 1 - \frac{2 M}{r} \right) }{S(r)^2} \, .
   \label{rdot2M}
  \ee
The $\dot{r}^2$ quantity as a function of $r$ is plotted in Fig.~\ref{rdotMSchw} for different values of the black hole mass $M$. For very small values of the mass $M$ with respect to the scale $L$, assuming $e=1$, there is a region between two positive values of the radial coordinate where the purely radial motion is impossible to be realized because $\dot{r}^2<0$. First, this classically forbidden region occurs always for radial coordinate bigger than the location of the horizon. If
\be
L < \frac{\sqrt{51 \sqrt{17}-107}1}{4} \, M \approx 2.54 \, M \, ,
\ee
then there is not such a region for $e=1$. If $L \gtrsim 2.54 \, M$, then the region bounded by two radii exists. We emphasize that its innermost boundary is still before the Schwarzschild horizon. When the black hole mass $M$ goes to zero, or the parameter $L$ tends to infinity, this excluded region is everywhere outside the horizon. If the energy of the particle is raised above $e=1$, then the limits of this region shrinks and the black hole masses which satisfy the condition must be even smaller. However, it is always possible to find a positive, bigger value of the energy $e$ that makes $\dot{r}^2$ positive again and everywhere. This is clear if we look at the effective potential $V_{\rm eff}$ in Fig.~\ref{rdotMSchw2}: contrary to the potential for the Schwarzschild metric, a peak at short distances shows up. Therefore, the spacetime is geodesically complete.

\begin{figure}
\begin{center}
\includegraphics[type=pdf,ext=.pdf,read=.pdf,height=5.0cm]{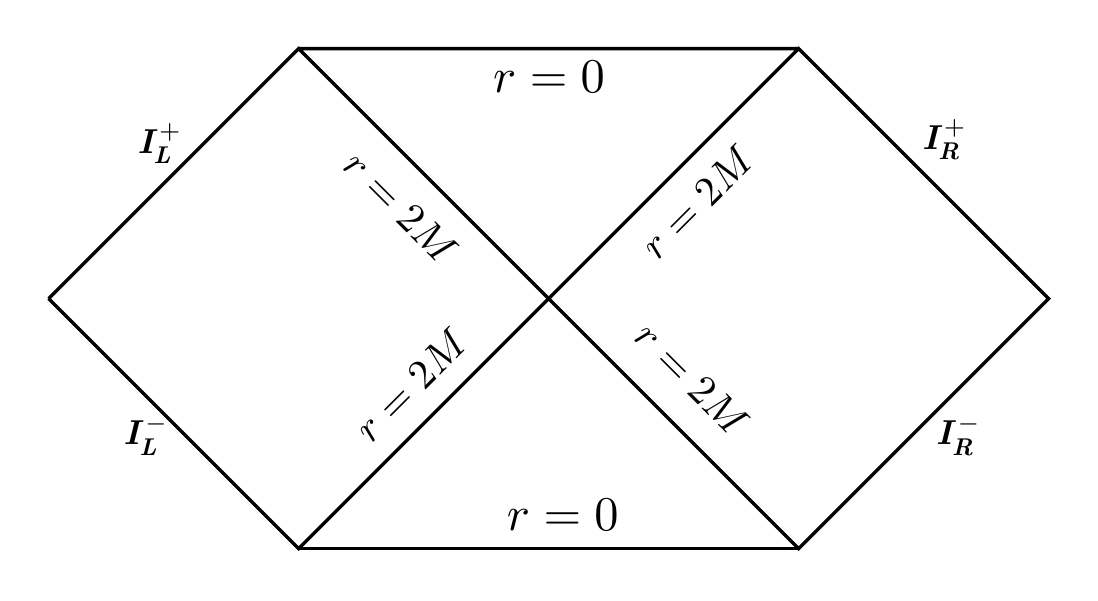}
\end{center}
\caption{Spacetime structure of the Schwarzschild metric in conformal gravity. This diagram has been derived changing coordinates to Kruskal-Szekeres ones. The overall  conformal factor $S(r)$ does not change the diagram and all the curves $r={\rm const.}$, $t={\rm const.}$, including $r=0$, are located in exactly the same positions as in the well known Schwarzschild diagram. However, now the spacetime is regular in $r=0$ and the horizontal line there can never be reached in a finite amount of proper time by any massive particle conformally or non-conformally coupled. Moreover, photons cannot reach $r=0$ for any finite value of the affine parameter $\lambda$. 
\label{PenroseD}}
\end{figure}

When we are very close to $r=0$, equation~\eqref{rdot2M} simplifies to 
  \be
  \dot{r}^2 
  \approx \frac{2 M }{L^4} \, r^3 \quad \Longrightarrow \quad
   \dot{r}   \approx - \frac{\sqrt{2 M } }{L^2} \, r^{3/2} \,  ,
  \label{EG0}
  \ee
for an infalling particle, where the metric is rescaled by the conformal factor $S(r)$ given by~(\ref{grazieMode}). Above we have assumed that the particle is falling in the black hole, hence the radial coordinate must decrease with time, $\dot r\leq0$, and this is the reason why the minus sign was chosen.

From $V_{\rm eff} = - \hat{g}_{tt}$, we infer that any massive particle can arrive at $r=0$. However, integrating equation~(\ref{EG0}), the proper time to reach the origin $r\to0^+$ turns out to be infinite 
\be
\Delta \tau \approx \frac{2 L^2}{\sqrt{2 M }} \left( \frac{1}{ \sqrt{r} } - \frac{1}{ \sqrt{r_0} } \right)  
\quad \Longrightarrow \quad 
\Delta \tau \equiv \tau(0^+) - \tau(r_0) \rightarrow +  \infty \, . 
\ee
The maximal extension of the black hole spacetime is given by the usual Penrose diagram in Fig.~\ref{PenroseD}. Neither massive particles nor photons (see later in this section) reach the point $r=0$ in a finite amount of time.

\begin{figure}
\begin{center}
\includegraphics[type=pdf,ext=.pdf,read=.pdf,height=7.0cm]{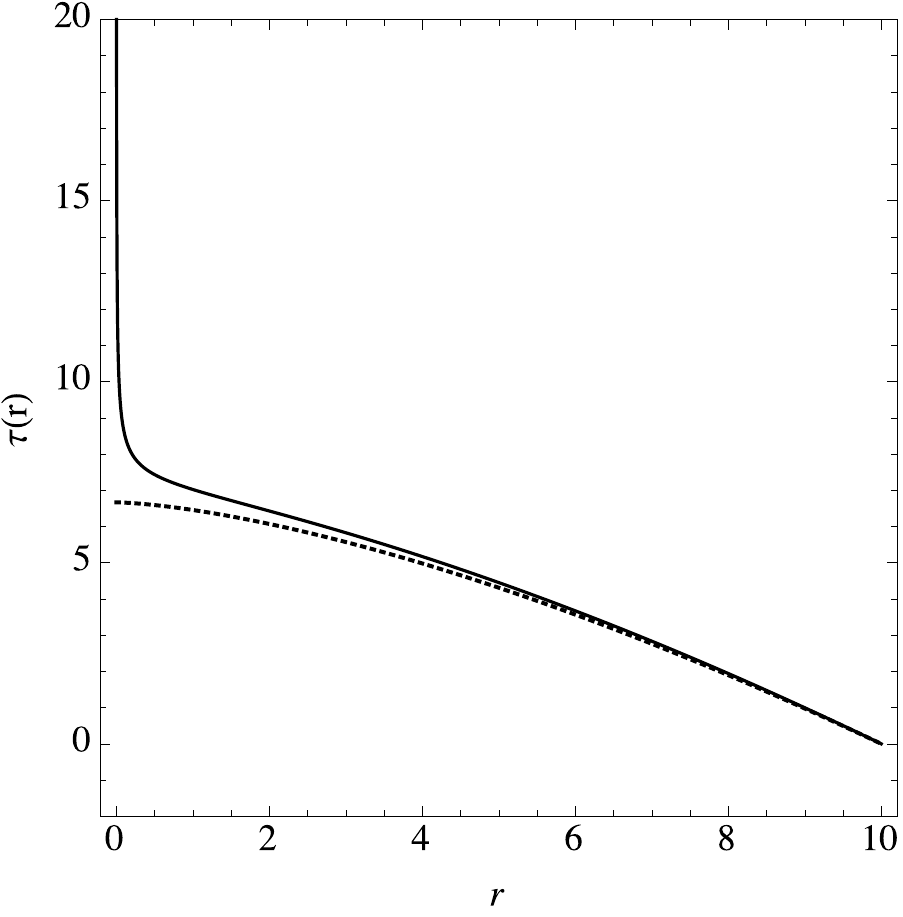}
\end{center}
\caption{Plot of the proper time $\tau$ needed for a massive particle to reach the point $r=0$ from the initial radial position $r_0$. The solid line corresponds to the radial geodesic in the metric rescaled by~(\ref{grazieMode}), while the dashed line is for the proper time in the Schwarzschild metric. Here $e=1$, $L=1$, $M=5$, and $r_0=10$. The result is qualitatively the same for any value of the constant $e$. \label{tauMssive}}
\end{figure}

We now integrate equation~(\ref{geometricabella2}) for some initial radial position $r_0$ (we are considering the case $\dot{r} <0$)
 \be
 \frac{S(r) | \dot{r}| }{\sqrt{e^2 - S(r) \left(1 - \frac{2 M}{r} \right) }} = 1 \, \quad 
 \Longrightarrow \quad 
 \tau = - \int_{r_0}^r \frac{S(r) dr}{\sqrt{e^2 - S(r) \left(1 - \frac{2 M}{r} \right) }}  \, .
 \label{taumexact}
 \ee
 For $L=0$, the exact solution for the Schwarzschild metric reads ($e = 1$)
 \be
 \tau =  \frac{2 \, r_0^{3/2}}{3 \sqrt{2 M}} - \frac{2 \,  r^{3/2}}{3 \sqrt{2 M}} \, .
 \label{tauSchm}
 \ee
In the Schwarzschild background, a test-particle reaches the singularity at $r=0$ in a finite proper time. The integral~(\ref{taumexact}) can be evaluated numerically and the solution is plotted in Fig.~\ref{tauMssive} (solid line) together with the Schwarzschild case~(\ref{tauSchm}) (dashed line). Clearly, in the spacetime described by the conformally rescaled metric, the massive particle never reaches $r=0$. Such a conclusion is qualitatively independent of the value of $e$. Indeed, even faster probes, possessing non-vanishing velocities at spatial infinity, needs infinite amount of proper time to reach the point $r=0$. Of course this is also true for particles on bounded orbits (with $e<1$) that are slower, at the corresponding radial locations, than particles on marginally bounded ($e=1$) orbits considered here.

The infinite amount of time needed to reach $r=0$ is a universal property common to all regular spacetimes obtained by applying a conformal analytic transformation to the Schwarzschild metric.

\begin{figure}
\begin{center}
\includegraphics[type=pdf,ext=.pdf,read=.pdf,height=7.0cm]{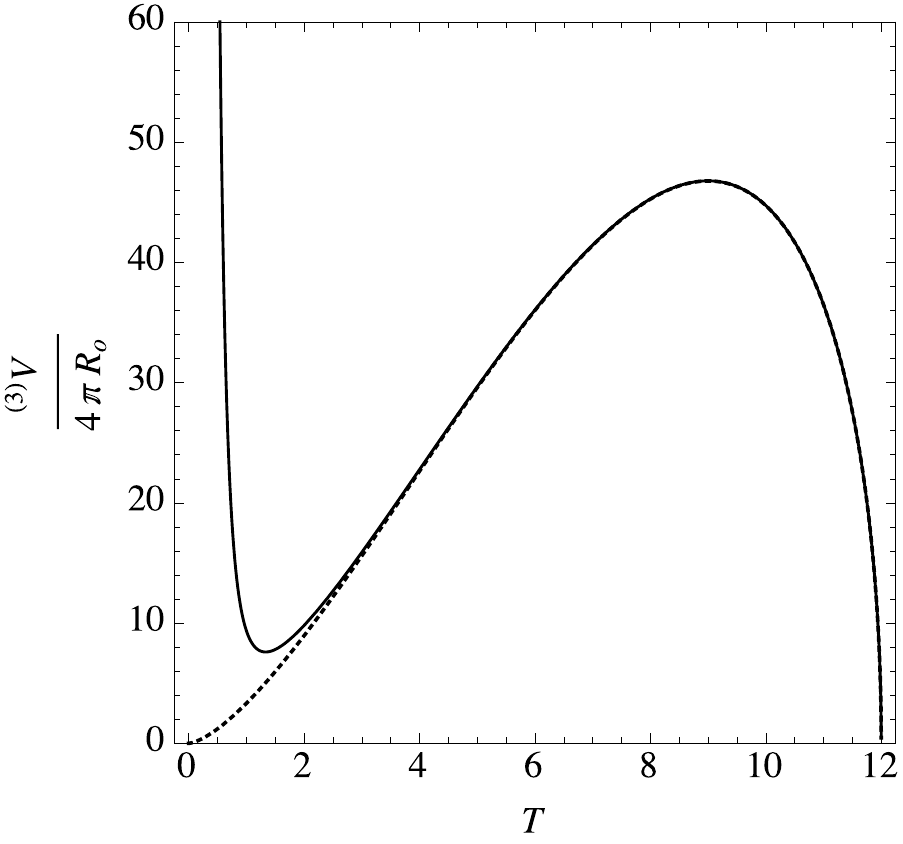}
\end{center}
\caption{Plot of the infrared renormalized three-volume $V^{(3)}/(4 \pi  R_o)$ given in~(\ref{3VSCH}). Here $L=1$ and $M=6$. The dashed line is the result for the Schwarzschild case. \label{Volume3}}
\end{figure}

Let us now evaluate the volume of the black hole interior, namely the volume inside the event horizon. For $r<2M$ the radial and time coordinates exchange their role, namely: $r=T$ and $t=R$. The metric belongs to the class of Kantowski-Sachs spacetimes, 
\be
d s^{*2} = S(T) \left[-  \frac{dT^2}{\frac{2 M}{T} -1}  + \left( \frac{2M}{T} -1 \right) d R^2 + T^2 d \Omega^{2} 
\right] \, , \quad T< 2M \, , 
\ee
and the interior spatial volume reads,
\be
V^{(3)} = 4 \pi  R_o  \,  S(T)^{3/2} \, T^2 \sqrt{\frac{2 M}{T} -1 } 
\ee
that for the choice of the conformal factor $S(r)$ as in~\eqref{grazieMode} turns in  
\be V^{(3)} =
4 \pi R_o  T^2 \sqrt{ \left(\frac{L^4}{T^4}+1\right)^3 \left(\frac{2 M}{T}-1\right)}
 \, , \quad T< 2M \,  ,
 \label{3VSCH}
\ee
where $R_o$ is an infrared cut-off due to the translational invariance in the radial variable $R$ of the metric inside the event horizon. The volume does not shrink to zero as in the Schwarzschild case, but reaches a minimum value and bounces back to infinity for $T \rightarrow 0$ (see Fig.~\ref{Volume3}.)

It deserves to be noticed that near $T = 0$, both the scale factors $g_{RR} \sim T^{-5}$ and $g_{\theta \theta} \sim T^{-2}$ undergo infinite expansion, but without isotropization.  A mechanism of isotropization due to particle creation has been proposed in~\cite{Bronnikov:2005gm} and can be applied here too. 
However, in our model there is no extension beyond $T=0$ because nobody and nothing can reach such a point (see also the next two sections about geodesic completion). Therefore, the isotropization mechanism should take place in the black hole interior itself, $0<T< 2M$. This is consistent with an infinite volume near $T=0$ that tuns in an isotropic Universe whether the mechanism mentioned above takes place.

\subsubsection{Conformally coupled test-particle probe}

\begin{figure}
\begin{center}
\includegraphics[type=pdf,ext=.pdf,read=.pdf,height=7.0cm]{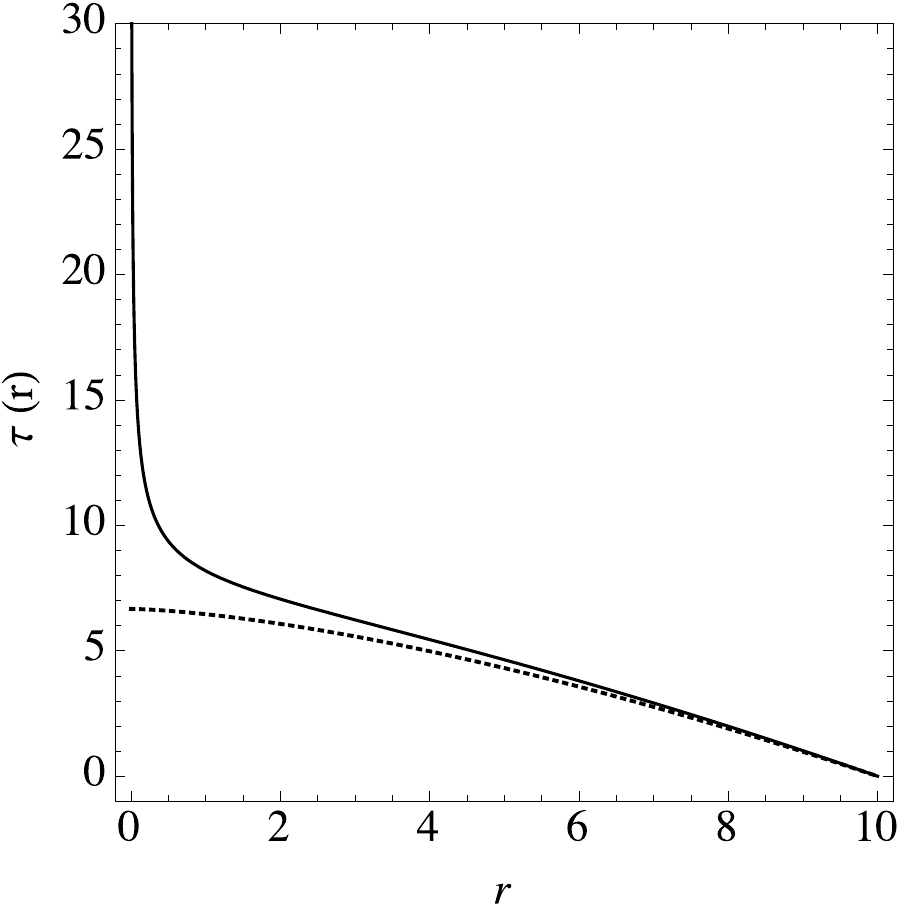}
\end{center}
\caption{Plot of the proper time as a function of the radial Schwarzschild coordinate for a particle falling into a black hole for the regular spacetime (solid line) and for the Schwarzschild spacetime (dashed line). Here $L=2$, to amplify the difference between the two lines, and $M=5$. The particle starts at $\tau=0$ from the horizon located at $r=2M$ (with non-zero velocity because it was at rest at spatial infinity).
 \label{TimeToSing}}
\end{figure}

In this section we study the geodesic completion by probing the spacetime with a test-particle conformally coupled to the Weyl-invariant gravitational theory. The  four-dimensional action is obtained by replacing again the metric $g_{\mu\nu}$ with $\phi^2 \kappa_4^2 \, \hat{g}_{\mu\nu}$~\cite{BekCP}, 
\be
S_{\rm cp} = - \int \sqrt{ - f^2 \phi^2 \hat{g}_{\mu\nu} d x^\mu d x^\nu} 
=  - \int \sqrt{ - f^2 \phi^2 \hat{g}_{\mu\nu} \frac{d x^\mu}{d \lambda} \frac{d x^\nu}{d \lambda} } 
\, d \lambda \, , 
\label{Spc}
\ee
where $f$ is a positive constant coupling strength, $\lambda$ is an affine parameter, and $x^\mu(\lambda)$ is the trajectory of the particle. In the unitary gauge $\phi = \kappa_4^{-1}$ the action~(\ref{Spc}) turns into the usual one for a particle with mass $m = f \kappa_4^{-1}$ ($f>0$). The Lagrangian reads 
\be
L_{\rm cp} = - \sqrt{ - f^2 \phi^2 \hat{g}_{\mu\nu} \dot{x}^\mu \dot{x}^\nu } \, , 
\ee
and the translation invariance in the time-like coordinate $t$ implies 
\be
\frac{\partial L_{\rm cp}}{\partial \dot{t} } = - \frac{f^2 \phi^2 \hat{g}_{tt} \dot{t}}{L_{\rm cp}} = {\rm const.} = - E \quad \Longrightarrow \quad \dot{t} = \frac{L_{\rm cp} E }{f^2 \phi^2 \hat{g}_{tt}} . 
\label{ConstE}
\ee
Since we are interested in evaluating the proper time of the particle necessary to reach the point $r=0$, we must choose the proper time gauge, namely $\lambda = \tau$. In this case, $E$ is the energy of the test-particle and
\be
\frac{d \hat{s}^2}{d \tau^2} = - 1 
\quad \Longrightarrow \quad  L_{\rm cp}= - f \phi 
\quad \Longrightarrow \quad  \dot{t} = - \frac{ E }{f \phi \, \hat{g}_{tt}} \, .
\label{PTG}
\ee
Replacing~(\ref{ConstE}) in $\hat{g}_{\mu\nu} \dot{x}^\mu \dot{x}^\nu = - 1$ and using the solution of the EOM for $\phi$, namely $\phi = S^{-1/2} \kappa^{-1}_4$, we end up with 
\be
S(r)^2 \dot{r}^2 + S(r) \left( 1 - \frac{2M}{r} \right) - \frac{E^2 \kappa_4^2}{f^2} S(r) = 0 .
\ee
For a particle at rest at infinity $E = f \kappa^{-1}_4$ and the above equation simplifies to 
\be
S(r) \dot{r}^2 = \frac{2M}{r} \, .
\label{Es}
\ee
For the scale factor~(\ref{geoS}), we can easily integrate~(\ref{Es}). The proper time to reach a general radial position $r$ starting from the position at the event horizon at $r=2 M$ reads
\be
\tau = \frac{4 M^2-3 L^2}{3 M}-\frac{\left(r^2-3 L^2\right) \sqrt{\frac{2 r}{M}}}{3 r} \, .
\ee
Note that for any value of $L\neq 0$ the particle never reaches the point  $r=0$, while for $L=0$ we recover the result of the finite amount of proper time that any particle needs to reach the singularity in the Schwarzschild metric, namely $\tau_{\rm Schw} = 4 M/3$ (see Fig.~\ref{TimeToSing}.)

\subsubsection{Geodesics for light rays}\label{luceSch}

\begin{figure}
\begin{center}
\includegraphics[type=pdf,ext=.pdf,read=.pdf,height=7.0cm]{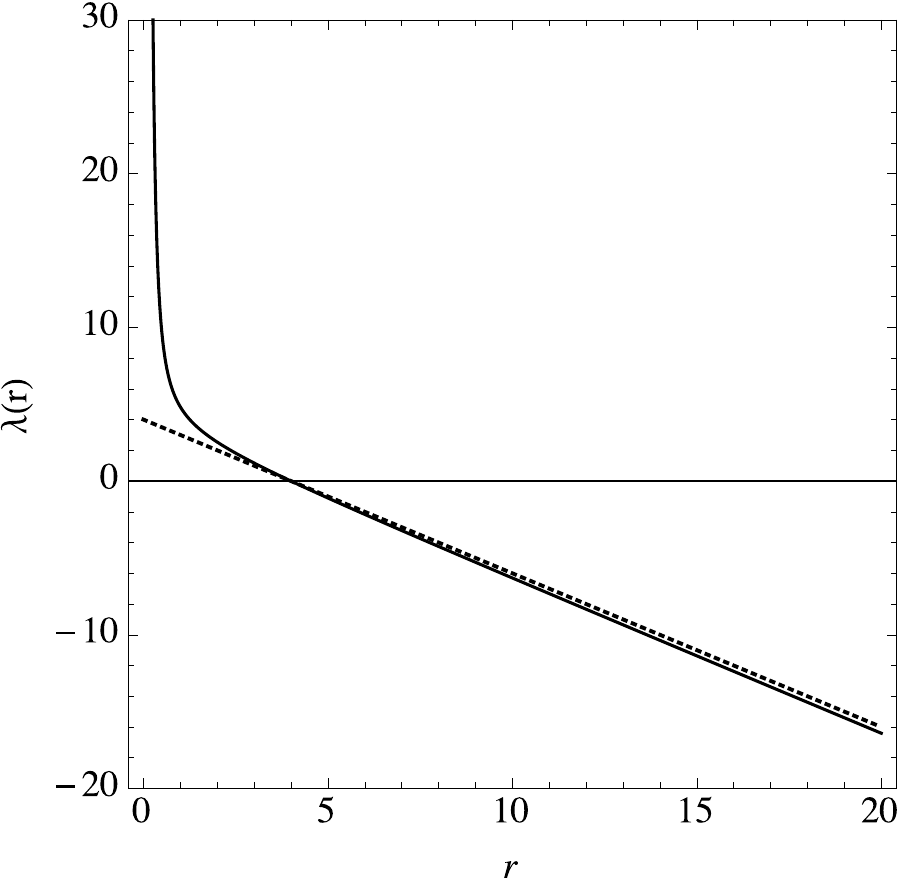}
\end{center}
\caption{Plot of the affine parameter $\lambda(r)$ for null geodesics in the rescaled singularity-free Schwarzschild metric (solid line) and in the Schwarzschild metric (dashed line). We here used the following values for the parameters and the conserved quantities: $M=1$, $L=1$, $r_0=4$, and $e=1$. \label{TimeToSingP}}
\end{figure}

In this section we derive and solve the radial geodesic equations for a massless particle in the singularity-free black hole spacetime~(\ref{NRBH}) with the conformal factor~(\ref{geoS}). The metric~(\ref{NRBH}) is time-independent and spherically symmetric (in particular it is invariant under $\varphi \rightarrow \varphi + \delta \varphi$). Therefore, we have the following Killing vectors associated with the above symmetries
\be
\xi^{\alpha} = (1, 0, 0, 0) \, , \quad \eta^{\alpha} = (0, 0, 0, 1) \, .
\ee
Since the metric is independent of the $t$- and $\varphi$-coordinates, we can construct the following conserved quantities\footnote{
To prove that the quantities $e$ and $\ell$ are conserved we have to use the geodesic equations of motion for massive, conformally coupled, or massless particles. It can be proved that the geodesic equations for light in the metric $\hat{g}_{\mu\nu}$ are independent on $\phi$ \cite{Mannheim2}. This is the reason why (\ref{Ldott}) and (\ref{Ldotphi}) do not depend on the dilaton $\phi$. 
}:
\be
&& e = - \xi \cdot u = - \xi^\alpha u^\beta \hat{g}_{\alpha \beta} = - \hat{g}_{ t \beta} u^\beta = - \hat{g}_{ t t } u^t = 
S(r) \left(  1 - \frac{2 M}{r} \right) \frac{d t}{d \lambda} = S(r) \left(  1 - \frac{2 M}{r} \right) \dot{t} \, , 
\label{Ldott} \\ 
&& \ell = \eta \cdot u = \eta^\alpha u^\beta \hat{g}_{\alpha \beta} =  \hat{g}_{ \phi \beta} u^\beta =  \hat{g}_{ \phi \phi } u^\phi = 
S(r) r^2 \sin^2 \theta \, \dot{\varphi} \, ,
\label{Ldotphi}
\ee 
where the null vector 
\be
u^{\alpha} = \frac{d x^\alpha}{d \lambda} 
\ee
satisfies 
\be
u \cdot u = \hat{g}_{\alpha \beta}  \frac{d x^\alpha}{d \lambda}  \frac{d x^\beta}{d \lambda} = 0 \, .
\label{NullV}
\ee
From~(\ref{NullV}), we get the following equation
\be
- \left(  1 - \frac{2 M}{r} \right) \dot{t}^2 + \frac{\dot{r}^2}{\left(  1 - \frac{2 M}{r} \right)} 
+ r^2 \sin^2 \theta \dot{\varphi}^2 = 0 \, .
\label{uu}
\ee
Note that the rescaling of the metric cancels out in the above equation \eqref{uu} for null geodesics, but $S(r)$ will appear again when the conserved quantities (\ref{Ldott}) and (\ref{Ldotphi}) are taken into account. Let us solve (\ref{Ldott}) for $\dot{t}$ and (\ref{Ldotphi}) for $\dot{\varphi}$ and, afterwards, replace the results in (\ref{uu}). The outcome is: 
\be
-  \frac{e^2}{S(r)^2 \left(1 - \frac{2M}{r} \right)} 
+ \frac{\dot{r}^2}{  1 - \frac{2 M}{r} } 
+ \frac{\ell^2}{ S(r)^2 r^2 } = 0 \, .
\label{uu2}
\ee
Let us focus on the radial geodesics (i.e. $\ell=0$), which will be sufficient to verify the geodesic completeness. Equation~(\ref{uu2}) simplifies to 
\be
-  \frac{e^2}{S(r)^2 } 
+\dot{r}^2 =0 \, \quad \Longrightarrow \quad S(r) | \dot{r} | = e \, .
\ee
The above first order differential equation can be easily integrated for a photon trajectory approaching $r=0$, namely for $\dot{r}=-\left(1-\frac{2M}{r}\right)^2\dot{t}<0$. 
\be
\lambda(r) = \frac{1}{e} \left[ \frac{L^4}{3 r^3}-\frac{L^4}{3 {r_0}^3}+\frac{2 L^2}{r}-\frac{2
   L^2}{{r_0}}-r+{r_0} \right]  . 
\ee
It turns out that photons cannot reach $r=0$ for any finite value of the affine parameter $\lambda$, as it is evident from Fig.~\ref{TimeToSingP}. Therefore, neither massive particle (including massive conformally coupled particle) nor photons can reach the point $r=0$ in a finite amount of time or other affine parameter of their geodesics.

\section{Avoiding the Kerr singularity in conformal gravity}

In this section, we provide a class of singularity-free rotating black hole solutions obtained by a conformal rescaling of the metric describing the Kerr geometry. As explained in Section~\ref{sec-2}, the rescaled metric is still an exact solution of the EOM of the theory, but corresponding to a non-trivial dilaton profile.

\subsection{Non-singular axially symmetric black hole}

The new singularity-free spinning black hole in $D=4$ (generalization to any dimension is straightforward) reads 
\be
&& ds^{* 2} \equiv 
\hat{g}_{\mu \nu}^* dx^\mu dx^\nu = S(r, \theta) \hat{g}_{\mu\nu}^{\rm Kerr} dx^\mu dx^\nu
\label{SKerr} \\
&& \phi^* = S(r, \theta)^{-1/2} \kappa_4^{-1} \, , 
\label{phi51}
\ee
where the selected scale factor $\Omega^2$ now depends on both the radial and angular coordinates, namely
\be
S(r) = \left( 1+ \frac{L^2}{\rho^2} \right)^4.
\label{grazieModeK}
\ee
We also remind the Kerr metric in Boyer-Lindquist coordinates:
\be
&& ds^2_{\rm Kerr} = - \left( 1 - \frac{r_s r}{\rho^2} \right) dt^2 + \frac{\rho^2}{\Delta} dr^2 -  
\frac{2 r_s r \, a}{\rho^2} \sin^2 \theta \, dt \, d \varphi + \rho^2 d \theta^2 
+ \left( r^2 + a^2 + \frac{r r_s a^2 \sin^2 \theta }{\rho^2} \right) \sin^2 \theta d \varphi^2 \, . \\
&& \rho^2 = r^2 + a^2 \cos^2 \theta \, , \quad \Delta = r^2 - r_s r + a^2 \, , \quad r_s  = 2 M \,, \quad  a = a^* M \, .
\label{Kerrr}
\ee
Above $(r,\theta,\phi)$ are standard spherical coordinates and the spin parameter $a$ takes the values from $0$ to $M$, while $a^*$ from $0$ to $1$.

The expression for the Ricci scalar of conformally rescaled Kerr metric by the factor $S(r)$ in \eqref{grazieModeK}  reads
\begin{eqnarray}
\hat R&=&- \frac{24 L^2 \left(a^2 x^2+r^2\right)}{\left(a^2 x^2+L^2+r^2\right)^{6} } \left[ a^6 x^6-2 a^6 x^4+3 a^4 L^2 x^4-4 a^4 L^2 x^2-4 a^4 M r x^4+a^4 r^2 x^4\right.\nonumber\\
   &&\left.-4a^4 r^2 x^2-4 a^2 L^2 M r x^2-4 a^2 L^2 r^2-a^2 r^4 x^2-2 a^2 r^4+8 L^2 M r^3-3 L^2 r^4+4 M r^5-r^6\right]\,,
\end{eqnarray}
where $x=\cos\theta$ \footnote{ 
We use $x$ to denote the angular variable $\cos \theta$ only for writing expressions of curvature invariants in the Kerr metric, in Fig.\ref{KK}, Fig.\ref{KKLargeastar}, and in the Appendix \ref{ap-1}. Later in the main text we will use $x$ to label the position of a test particle.}. One notices that this expression is everywhere regular.
We have also evaluated the Kretschmann invariant for the metric (\ref{SKerr}), but to avoid a cumbersome formula in the text its expression is reported in Appendix~\ref{ap-1}. The plot of the the Kretschmann invariants for small and large $a^*$ are shown in Fig.~\ref{KK} and 
Fig.~\ref{KKLargeastar} respectively.


\begin{figure}
\begin{center}
\includegraphics[type=pdf,ext=.pdf,read=.pdf,height=7.5cm]{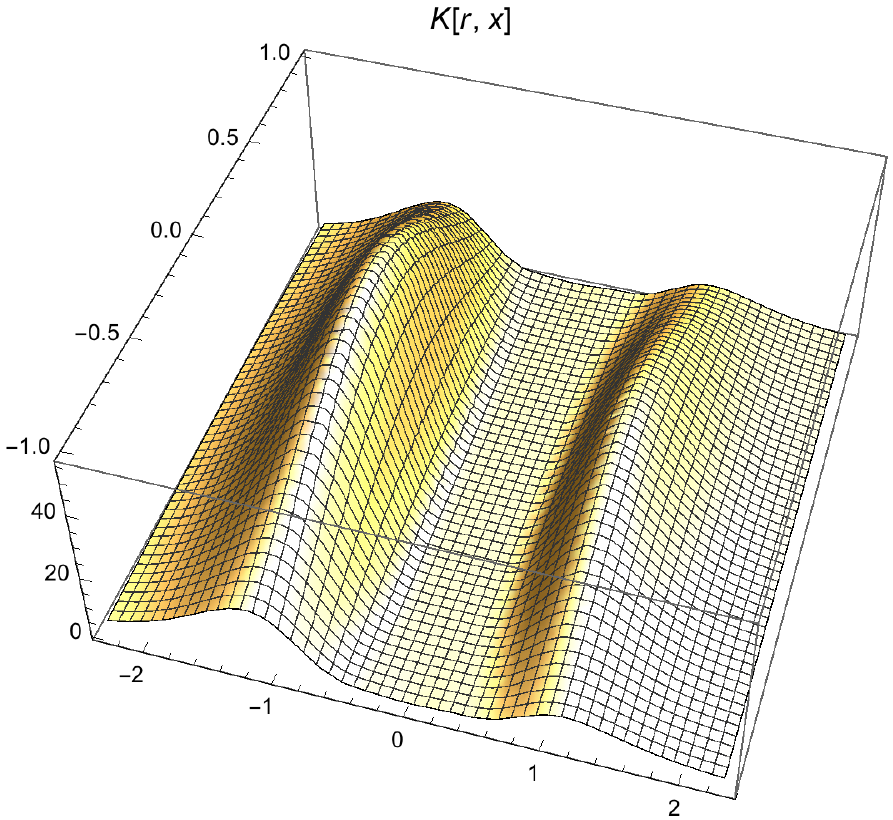}
\hspace{1.5cm}
\includegraphics[type=pdf,ext=.pdf,read=.pdf,height=7.5cm]{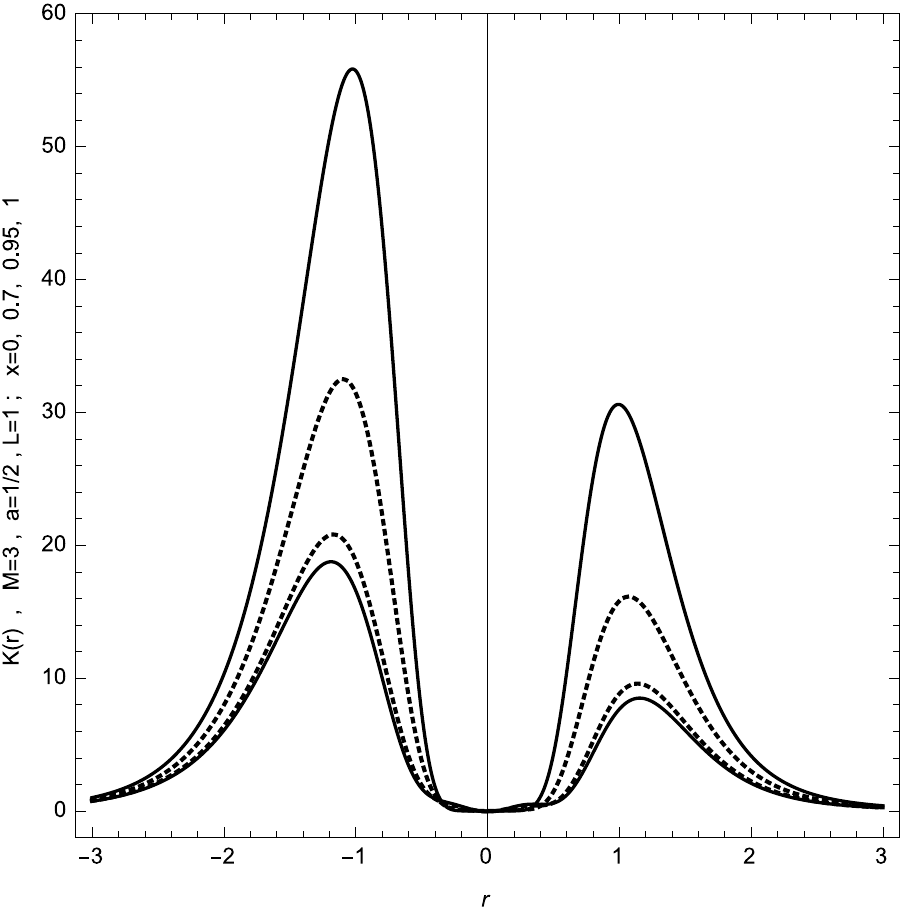}
\end{center}
\caption{Panel on the left: Kretschmann invariant for the singularity-free Kerr metric on the plane $(r,x=\cos\theta)$ (we here used $M=3$, $a^*=1/6$ 
, $L=1$). 
Panel on the right: Kretschmann invariants for four different fixed values of the $x$ parameter:
$x=1, 0.95, 0.7$ and  $0$, which correspond respectively to the curves from the bottom to the top in the panel. 
The physical reason to introduce these plots is to show not only the finiteness, but also the smoothness of the Kretchmann in contrast to other regular axi-symmetric spacetimes in the literature \cite{Spallucci, CaravelliModesto, ModestoNicolini, Noi2,DeLorenzo:2015taa}.
\label{KK}}
\end{figure}


  \begin{figure}
\begin{center}
\includegraphics[type=pdf,ext=.pdf,read=.pdf,height=7.5cm]{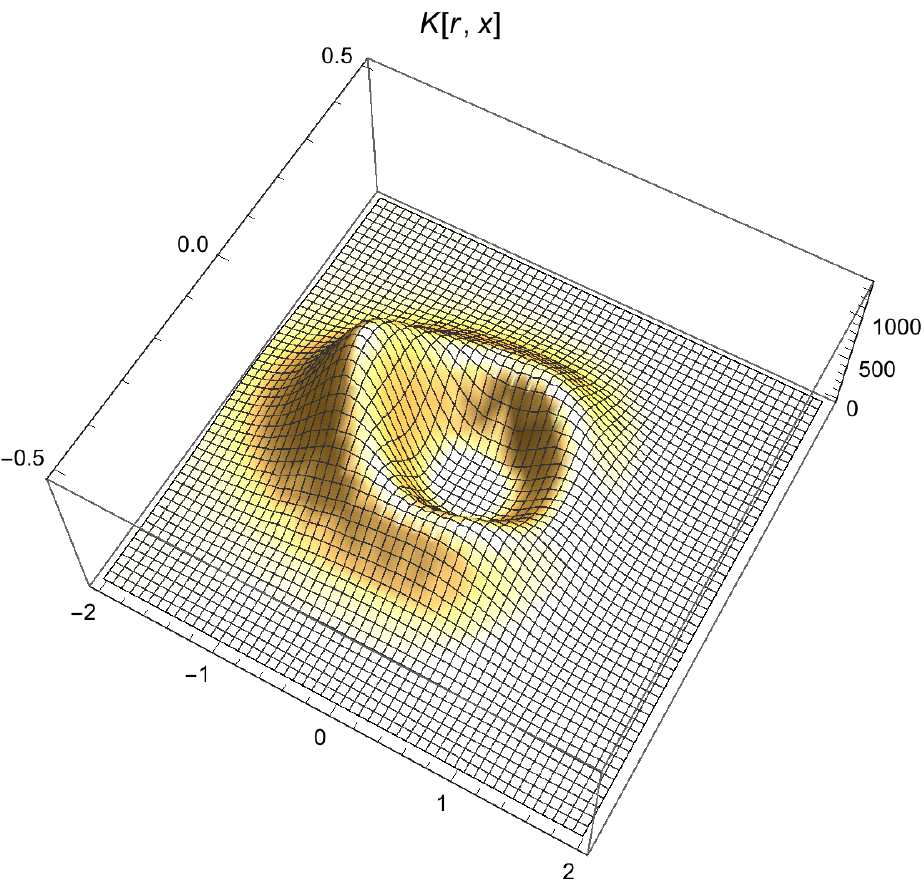}
\hspace{1.5cm}
\includegraphics[type=pdf,ext=.pdf,read=.pdf,height=7.5cm]{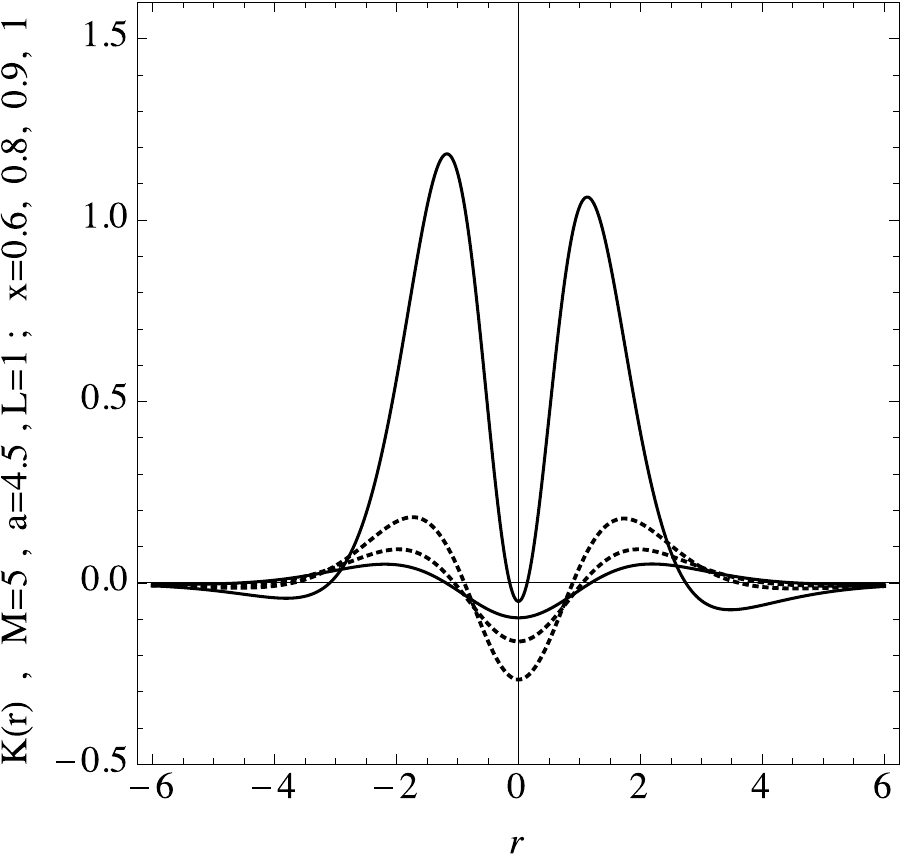}
\end{center}
\caption{Panel on the left: Kretschmann invariant for the singularity-free Kerr metric on the plane $(r,x=\cos\theta)$ (we here used $M=5$, $a^*=0.9$ 
, $L=1$). 
Panel on the right: Kretschmann invariants for four different fixed values of the $x$ parameter:
$x=1, 0.95, 0.7$ and  $0$, which correspond respectively to the curves from the bottom to the top in the panel. Here $M=5$ and $a^*=0.9$
\label{KKLargeastar}}
\end{figure}

\begin{figure}
\begin{center}
\hspace{-0.6cm}
\includegraphics[type=pdf,ext=.pdf,read=.pdf,height=5.0cm]{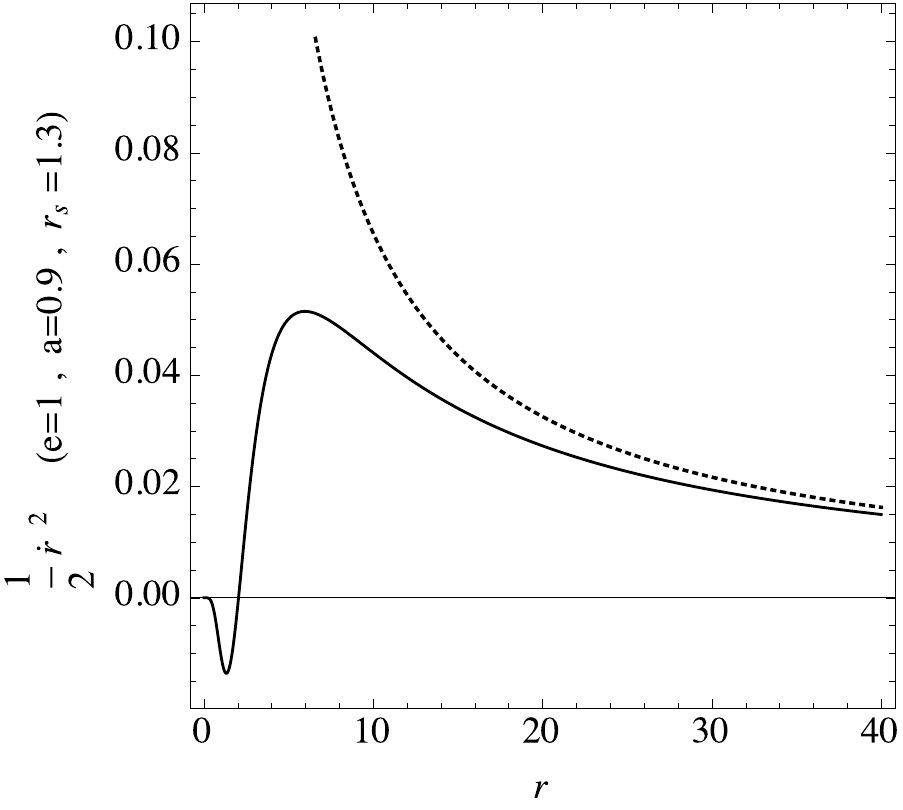}
\includegraphics[type=pdf,ext=.pdf,read=.pdf,height=5.0cm]{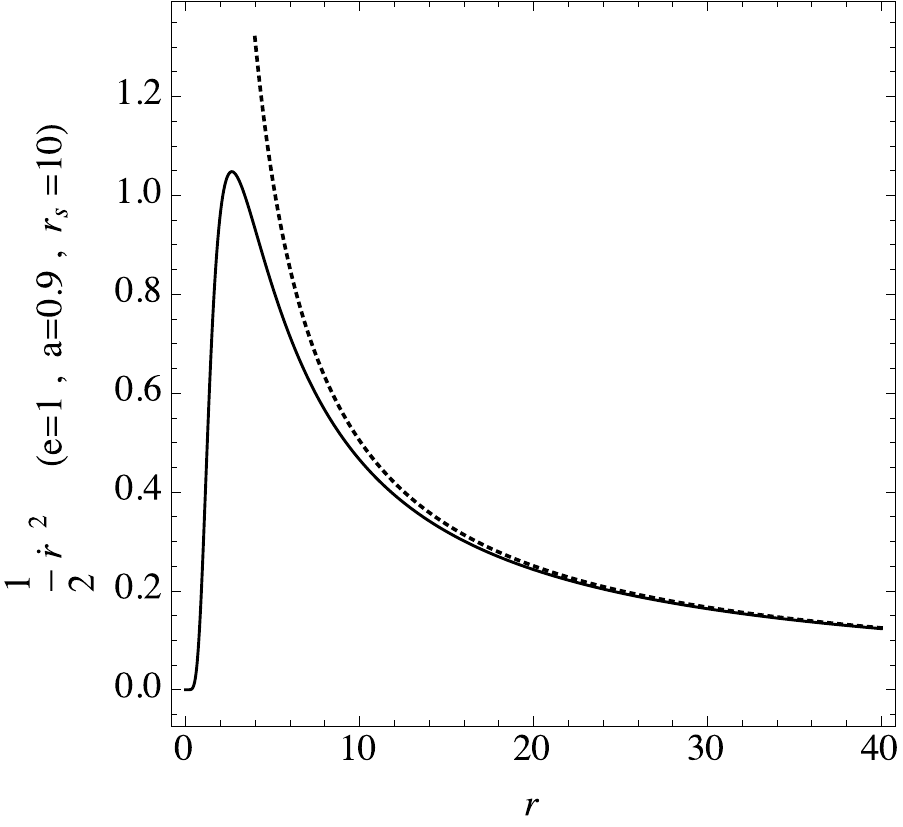}
\includegraphics[type=pdf,ext=.pdf,read=.pdf,height=5.0cm]{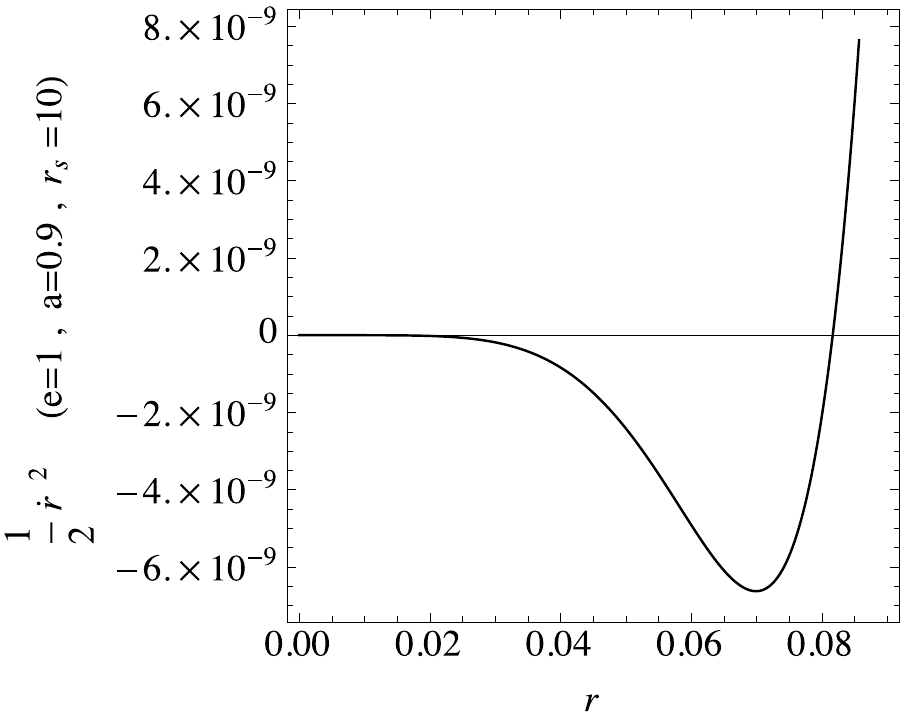}
\end{center}
\caption{Plots of $\dot{r}^2/2$ for $r_s=1.3$ and $r_s=10$, and $e=1$, $a=0.9$, $\theta=\pi/2$ and $L=1$ (solid line) or $L=0$ (dashed line - Kerr metric). The third plot for $r_s=10$ highlights the negativity of the ``kinetic energy" for $r<\tilde{r}$ regardless of the value of the mass. \label{VeffKerr}}
\end{figure}

\begin{figure}
\begin{center}
\includegraphics[type=pdf,ext=.pdf,read=.pdf,height=5.0cm]{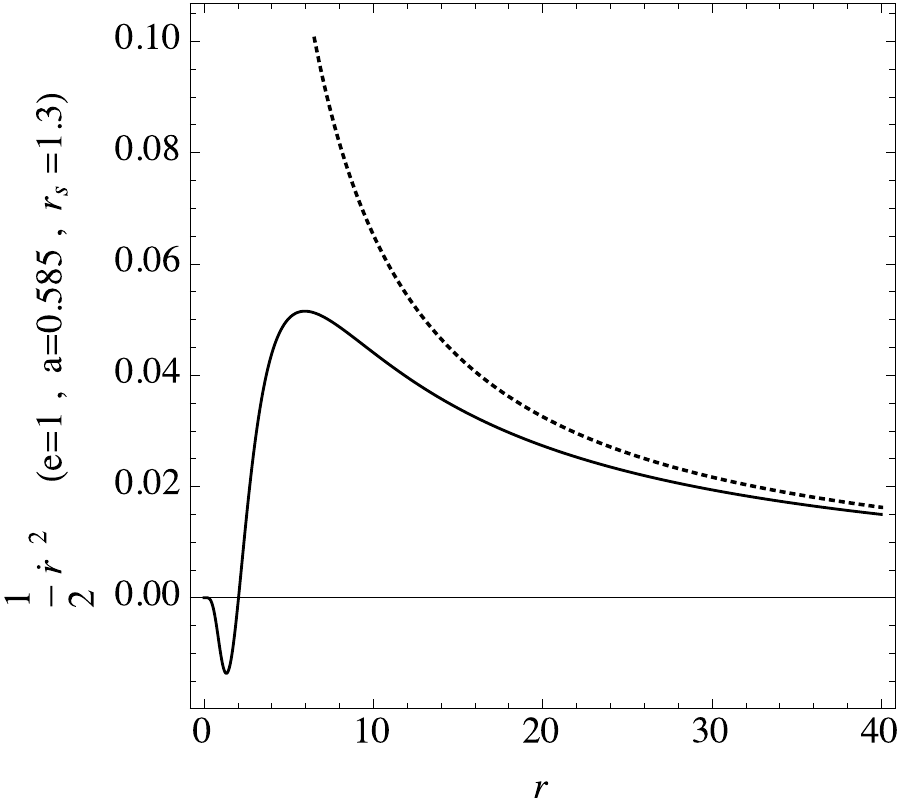}
\includegraphics[type=pdf,ext=.pdf,read=.pdf,height=5.0cm]{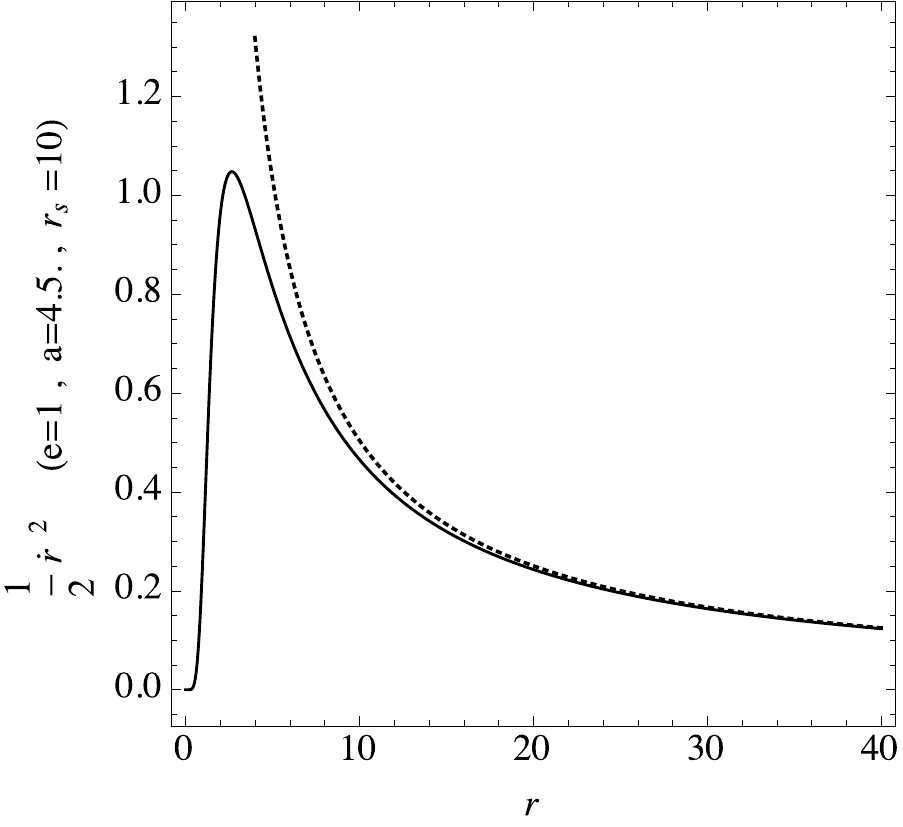}
\includegraphics[type=pdf,ext=.pdf,read=.pdf,height=5.0cm]{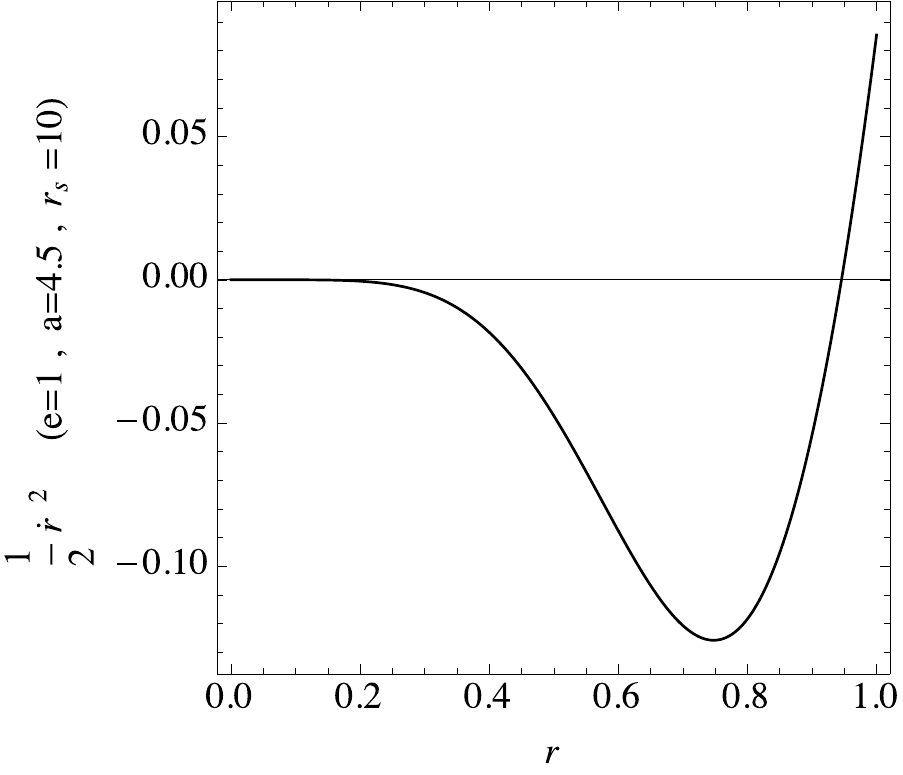}
\end{center}
\caption{Plots of $\dot{r}^2/2$. In the first plot from the left $r_s=1.3$, $a=0.585$, $e=1$, $\theta=\pi/2$, and $L=1$ (solid line) or $L=0$ (dashed line - Kerr metric). In the second plot $r_s=10$ and $a=4.5$. The third plot is for $r_s=10$ and $a=4.5$ or $a^*=0.9$. \label{VeffKerrLargeastar}}
\end{figure}

It deserves to be noticed that the curvature invariant does not suffer from any discontinuity when the classical singularity is approached from $\theta = \pi/2$ or $\theta\neq\pi/2$; see the discussion in~\cite{Noi2} for more details. This is a feature of the new class of solutions that is not shared with the other non-singular spacetimes constructed in the past \cite{Noi2,Spallucci,CaravelliModesto, ModestoNicolini}. Moreover, the exact solutions presented in this section can be derived applying the Newman-Janis (NJ) algorithm to the non-singular Schwarzschild solution extensively studied in the previous section.

\subsection{Geodesic completion}

\subsubsection{Geodesic completion based on the motion of massive particles} 

In this subsection we compute the radial geodesics for conformally and non-conformally coupled particles moving in the equatorial plane of the non-singular conformally rescaled Kerr geometry. Let us start explicitly with writing of the Lagrangian for a massive particle moving in the equatorial plane ($\theta=\pi/2$) of a general axi-symmetric spacetime with the metric in the canonical form
\be
\mathcal{L}_m = - m \sqrt{- \dot{x}^2} = - m \sqrt{-\left( \hat{g}_{tt} \dot{t}^2 + \hat{g}_{rr} \dot{r}^2 + 2 \hat{g}_{\varphi t} \dot{\varphi} \dot{t} + 
\hat{g}_{\varphi \varphi} \dot{\varphi}^2 \right)}\,,
\ee
where $m$ is the mass of the test particle. The invariance with respect to time translation and the invariance with respect to translation of the coordinate $\varphi$ imply,  respectively, that the following two quantities are conserved 
\be
&& \frac{\partial \mathcal{L}_m}{\partial \dot{t}} = - \frac{m^2}{\mathcal{L}_m} \left( \hat{g}_{tt} \dot{t} + \hat{g}_{t \varphi} \dot{\varphi} \right) = -E \, , 
\label{Ek}
\\
&& \frac{\partial \mathcal{L}_m}{\partial \dot{\varphi}} =  - \frac{m^2}{\mathcal{L}_m} 
\left( \hat{g}_{t \varphi} \dot{t} + \hat{g}_{\varphi \varphi} \dot{\varphi} \right) =  \ell \, . 
\label{ellek}
\ee
In the proper time gauge $\dot{x}^2 = -1$. The Lagrangian simplifies on-shell to $\mathcal{L}_m = - m$, while $E$ is the energy of the test-particle, and 
\be
\dot{x}^2 = -1 \quad \Longrightarrow \quad 
 \hat{g}_{tt} \dot{t}^2 + \hat{g}_{rr} \dot{r}^2 + 2 \hat{g}_{\varphi t} \dot{\varphi} \dot{t} + 
\hat{g}_{\varphi \varphi} \dot{\varphi}^2 = -1 \, .
\label{xdotte}
\ee
Solving equations (\ref{ellek}) and (\ref{Ek}) respectively for $\dot{\varphi}$ and $\dot{t}$ with $\mathcal{L}_m = - m$ and $\ell=0$, we get the radial geodesic equation for a massive particle, namely 
\be 
\dot{\varphi} = - \frac{\hat{g}_{t \varphi}}{\hat{g}_{\varphi\varphi}} \dot{t} \, , \quad 
 \dot{t} 
 = - \frac{E}{m} \left( \frac{\hat{g}_{\varphi\varphi}}{\hat{g}_{tt} \hat{g}_{\varphi\varphi} - \hat{g}_{t \varphi}^2 } \right) \, .
\ee
Using the above equations, (\ref{xdotte}) turns into 
\be
\hat{g}_{rr} \dot{r}^2 + e^2 \left( \frac{\hat{g}_{\varphi \varphi}}{\hat{g}_{tt} \hat{g}_{\varphi\varphi} - \hat{g}_{t \phi}^2 } \right) = -1 \, . 
\ee
where $e = E/m$ as before. Replacing in the above equation the metric components for their values as in~(\ref{Kerrr}), eventually we find
\be
S(r) \left(  \frac{r^2}{r^2 - r_s r +a^2} \right) \dot{r}^2 - e^2 S(r)^{-1} \left(\frac{r^3+ a^2 r+ a^2 r_s }{r^3-r^2 r_s+ a ^2 r}\right) = - 1\, ,
\ee
that we can write in a more familiar form introducing the following effective potential
\be
&& V_{\rm eff} = \frac{1}{2} \left(e^2-1\right)-\frac{a^2 \left(e^2 (r+ r_s )-r S(r) \right) +
r^2   \left(e^2 r+S(r) ( r_s-r)\right)}{2 r^3 S(r)^2} \, , \\
&& \frac{1}{2} \left( \frac{d r}{d \tau} \right)^2+ V_{\rm eff}(r) = \frac{e^2 -1}{2} \, .
\ee
A short discussion about the corresponding equations for the radial motion with a general non-zero value of the angular momentum $\ell$ is in the Appendix \ref{ap-2}.

The potential near $r=0$ is positive for any value of $e \geqslant 1$, and it has the following series expansion there
\be
 V_{\rm eff} = \frac{1}{2} \left(e^2-1\right) + \frac{a^2 }{2 L^8} r^6 
 -\frac{r_s }{2 L^8} r^7 
 + \left(\frac{1}{2 L^8}  -\frac{2 a^2}{L^{10}}\right) r^8
 +\frac{2  r_s }{L^{10}} r^9 
 +  \left(\frac{5 a^2}{L^{12}}-\frac{2}{L^{10}}\right) r^{10} 
  + O(r^{11}) \, .
\ee
The  ``kinetic energy" $\frac{1}{2}\dot{r}^2$ becomes zero for a positive value of the radial coordinate which we denote by $\tilde{r}$ (see Fig.~\ref{VeffKerr} and Fig.~\ref{VeffKerrLargeastar}).

We remind that for radial motion the potential in the Kerr metric (limit $L=0$ of the conformally rescaled metric) reads
\be
V_{\rm eff} = 
- \frac{{r_s}}{2 r}
-\frac{a^2 \left(e^2-1\right)}{2
   r^2}
- \frac{a^2 e^2 {r_s}}{2 r^3} \, .
\ee

For a particle approaching the point $r=0$ from positive values of the radial coordinate, we have to solve the following first order differential equation
\be
\hspace{-0.3cm}
\frac{r^3 S(r)^2}{a^2 (  r + {r_s} -r S(r) )-r^2 (r (S(r)-1) - {r_s} S(r))} \,  \dot{r}^2 =  1   
   \,\, \Longrightarrow \,\,
\sqrt{\frac{r^2 \left(\frac{L^2}{r^2}+1\right)^4}{\left(r (r-{r_s}) +a^2\right)
   \left(\frac{r^7 \left(r^3 + a^2 (r + r_s ) \right)}{\left(L^2+r^2\right)^4
   \left(r^2  - r r_s +a^2\right)}-1\right)}} \,  | \dot{r}| =  1 
\label{functionS}
\, . 
\ee
We note that the function that multiplies $\dot{r}^2$ becomes zero for a positive value of the radial coordinate $\tilde{r}$. This value is smaller than $r_-$ (the inner horizon of the Kerr metric), where the function assumes a finite positive value. For radial coordinates smaller than $\tilde{r}$, the classical motion is impossible, because we would have $\dot{r}^2<0$, which is forbidden in classical physics (there is however the possibility of quantum tunneling to such regions).
Therefore, the massive particles can never reach $r=0$. The detailed numerical and analytic investigation shows that that the location of $\tilde{r}$ is always under the inner horizon of the Kerr metric, that is $\tilde{r}<r_-=M-\sqrt{M^2-a^2}$. The difference however is very small and tends to zero, where the parameter $a$ vanishes. For the maximal allowed value of the parameter $a=M$ this difference is of the order of $M(1-2M^4 L^{-4})$ for $L\gg M$. Again, it is interesting to ask which kind of interpretation can be given to the  turning point $\tilde{r}$ occurring just under the inner horizon.

\begin{figure}
\begin{center}
\includegraphics[type=pdf,ext=.pdf,read=.pdf,height=7.0cm]{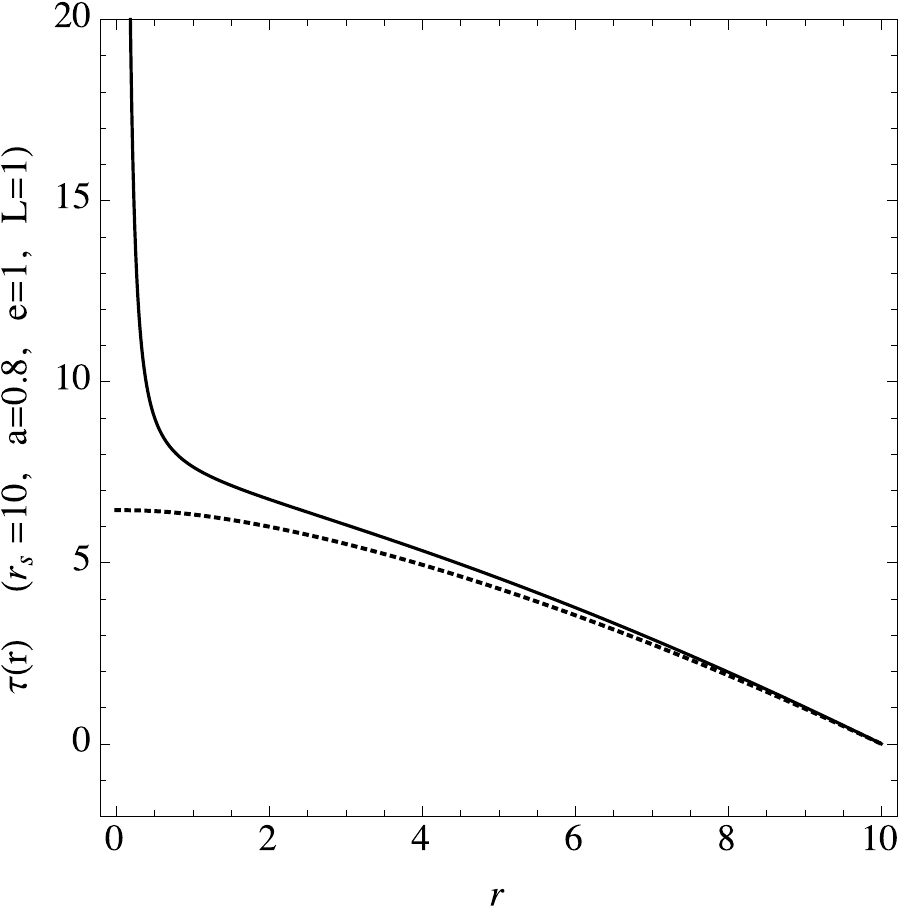}
\hspace{2cm}
\includegraphics[type=pdf,ext=.pdf,read=.pdf,height=7.0cm]{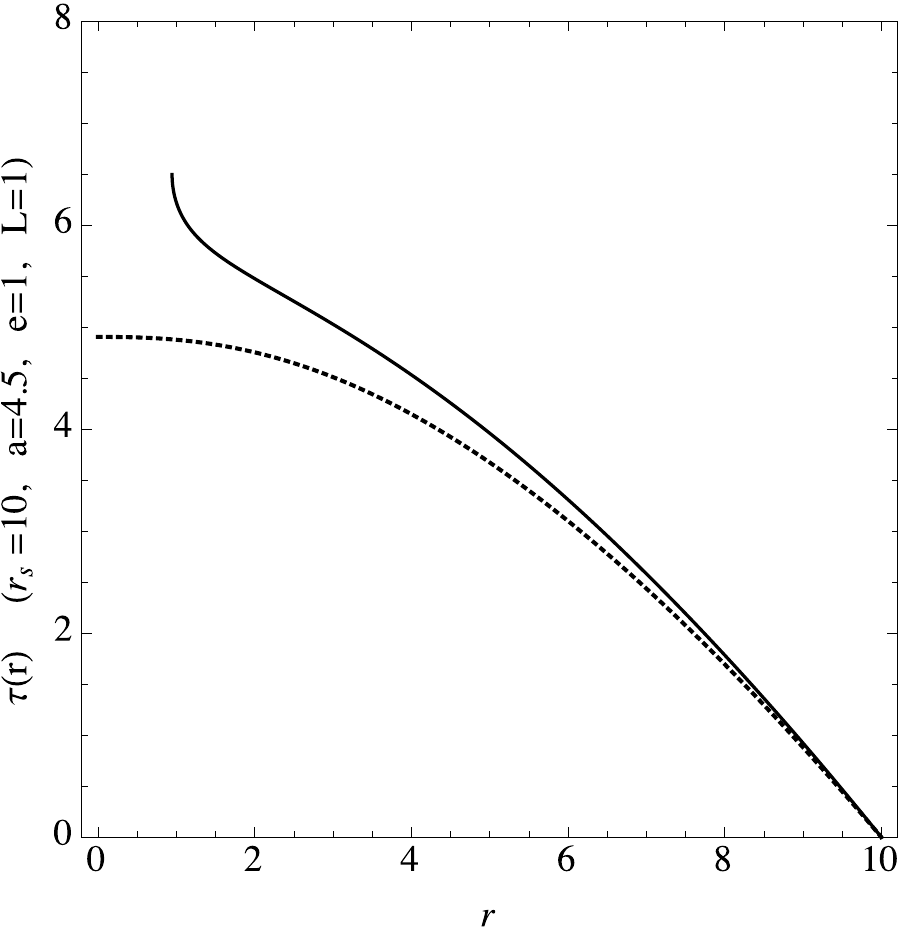}
\end{center}
\caption{Panel on the left: plot of the proper time $\tau(r)$ needed for a massive test-particle to reach the point $r=0$ from  the initial radial coordinate $r_0$ in the conformally rescaled Kerr metric (solid line) and in the Kerr metric (dashed line). Here we use the following values for the parameters and the conserved quantities: $r_s=10$, $L=1$, $r_0=10$, $e=1$, $\theta=\pi/2$ and $a=0.8$. Panel on the right: same physical quantity of the plot in the left panel, but for $a=4.5$.
\label{TimeToSingKM}}
\end{figure}

The proper time to reach $\tilde{r}$ is:
\be
\tau =  - {\int}_{r_0}^r \sqrt{\frac{r^2 \left(\frac{L^2}{r^2}+1\right)^4}{\left(a^2+r (r-{r_s})\right)
   \left(\frac{r^7 \left(a^2 (r+{r_s})+r^3\right)}{\left(L^2+r^2\right)^4
   \left(a^2+r (r-{r_s})\right)}-1\right)}} \, dr \, . 
\ee
The plot for the results of numerical integration is given in Fig.~\ref{TimeToSingKM}. The proper time to arrive to $\tilde{r}$ is finite as can be seen from the right panel in Fig.~\ref{TimeToSingKM}. The leading term of the expansion of the integrand near $\tilde{r}$ is $(r-\tilde{r})^{-1/2}$, hence the integral is convergent in the limit $r\to\tilde{r}$. This is brought about by the fact that the turning point is a single zero of the function in \eqref{functionS}, which multiplies $\dot{r}^2$. This again raises further interpretational issues.

\subsubsection{Geodesic completion for conformally coupled particles}

In this subsection we compute the radial geodesics for conformally coupled particles moving in the equatorial plane of the non-singular Kerr geometry. The Lagrangian for a conformally coupled particle is given in~(\ref{Spc}). In an axi-symmetric spacetime with the metric in canonical form, it simplifies to
\be
\mathcal{L}_{\rm cp} = -  \sqrt{- f^2 \phi^2  \dot{x}^2} = -  \sqrt{- f^2 \phi^2 \left( \hat{g}_{tt} \dot{t}^2 + \hat{g}_{rr} \dot{r}^2 + 2 \hat{g}_{\varphi t} \dot{\varphi} \dot{t} + 
\hat{g}_{\varphi \varphi} \dot{\varphi}^2 \right)} \, ,
\ee
where $\phi$ is the dilaton and $f$ is again a dimensionless coupling. Time translation invariance and rotation invariance in the angular coordinate $\varphi$ imply, respectively, that the following two quantities are conserved 
\be
&& \frac{\partial \mathcal{L}_{\rm cp}}{\partial \dot{t}} = - \frac{f^2 \phi^2}{\mathcal{L}_{\rm cp}} \left( \hat{g}_{tt} \dot{t} + \hat{g}_{t \varphi} \dot{\varphi} \right) = -E \, , 
\label{Ek2}
\\
&& \frac{\partial \mathcal{L}_{\rm cp}}{\partial \dot{\varphi}} =  - \frac{f^2 \phi^2}{\mathcal{L}_{\rm cp}} \left( \hat{g}_{t \varphi} \dot{t} + \hat{g}_{\varphi \varphi} \dot{\varphi} \right) =  \ell \, . 
\label{ellek2}
\ee
In the proper time gauge $\dot{x}^2 = -1$,  then
\be
\dot{x}^2 = -1 \quad \Longrightarrow \quad 
 \hat{g}_{tt} \dot{t}^2 + \hat{g}_{rr} \dot{r}^2 + 2 \hat{g}_{\varphi t} \dot{\varphi} \dot{t} + 
\hat{g}_{\varphi \varphi} \dot{\varphi}^2 = -1.
\label{xdotteCP}
\ee
Solving equations (\ref{ellek2}) and (\ref{Ek2}) respectively for $\dot{\varphi}$ and $\dot{t}$ with $\mathcal{L}_{\rm cp}= - f \phi$ and $\ell=0$, we get the radial geodesic equation for a massive particle, namely 
\be \dot{\varphi} = - \frac{\hat{g}_{t \varphi}}{\hat{g}_{\varphi\varphi}} \dot{t} \, , \quad 
 \dot{t} = - \frac{E}{f \phi} \left( \frac{\hat{g}_{\varphi\varphi}}{\hat{g}_{tt} \hat{g}_{\varphi\varphi} - \hat{g}_{t \varphi}^2 } \right) \, .
 \label{radialcp}
\ee
Using the above equations, (\ref{xdotteCP}) turns into 
\be
\hat{g}_{rr} \dot{r}^2 + \frac{E^2}{f^2 \phi^2} \left( \frac{\hat{g}_{\varphi\varphi}}{\hat{g}_{tt} \hat{g}_{\varphi\varphi} - \hat{g}_{t \varphi}^2 } \right) = -1 \, . 
\ee
We now replace the full solution~(\ref{SKerr}) in~(\ref{radialcp}), including the one for  the dilaton field $\phi = f \kappa_4^{-1} S^{-1/2}$. For the sake of simplicity, we again consider the case of a particle at rest at infinity, or $E= f \kappa_4^{-1}$. Finally, we get
\be
 S(r) \dot{r}^2  = \frac{r_s(a^2 +r^2)}{r^3} \, .
\ee
Note that now $\dot{r}^2$ is always positive for any $r>0$. For a particle approaching the point $r=0$ from positive values of the radial coordinate, we have to solve the following first order differential equation 
\be
| \dot{r}| =  \sqrt{\frac{r_s \left(a^2+r^2\right)}{r^3}} \frac{1}{ \left( 1+ \frac{L^2}{r^2} \right)^2}   \quad 
\Longrightarrow \quad 
\dot{r} =  - \sqrt{\frac{r_s r^5\left(a^2+r^2\right)}{(r^2 +L^2)^4}} \, . 
\ee
The proper time to reach $r=0$ is:
\be
\tau =  - \int_{r_0}^r \sqrt{\frac{(r^2 +L^2)^4}{r_s r^5\left(a^2+r^2\right)}} dr \, . 
\ee
The plot of $\tau(r)$ is very similar to that of $\tau(r)$ for a massive test-particle and is not shown here. The proper time to arrive at $r=0$ is infinite.

\subsubsection{Geodesic completion for photons}

\begin{figure}
\begin{center}
\includegraphics[type=pdf,ext=.pdf,read=.pdf,height=7.0cm]{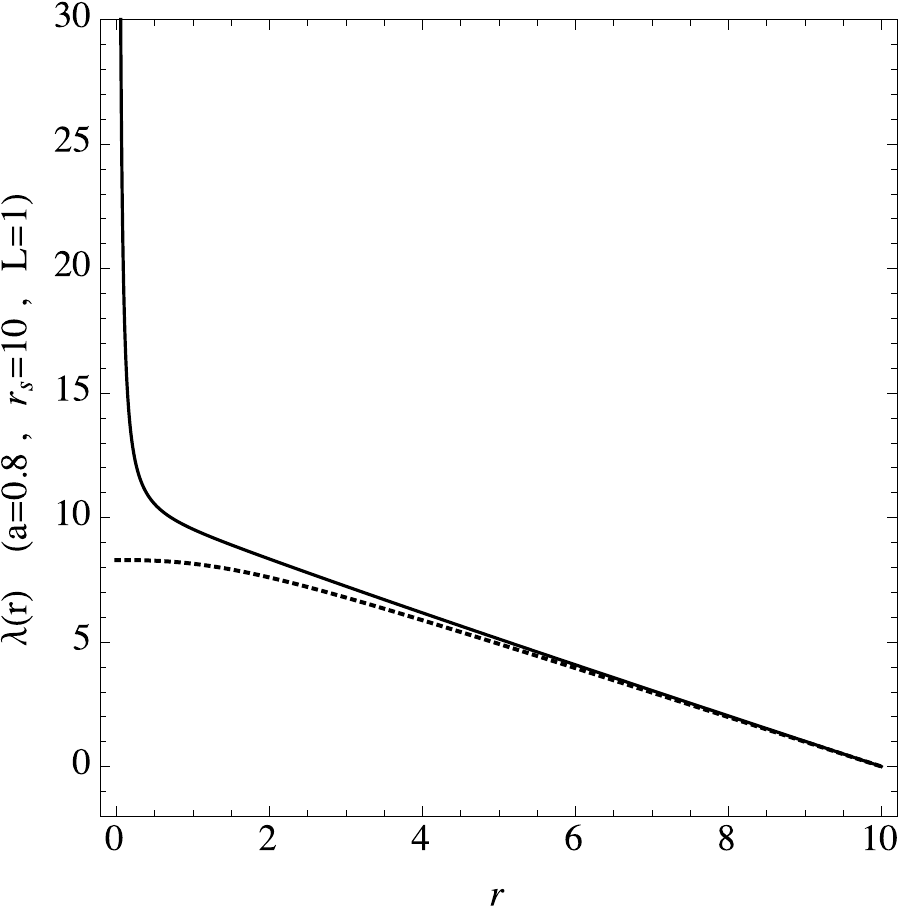}
\hspace{2cm}
\includegraphics[type=pdf,ext=.pdf,read=.pdf,height=7.0cm]{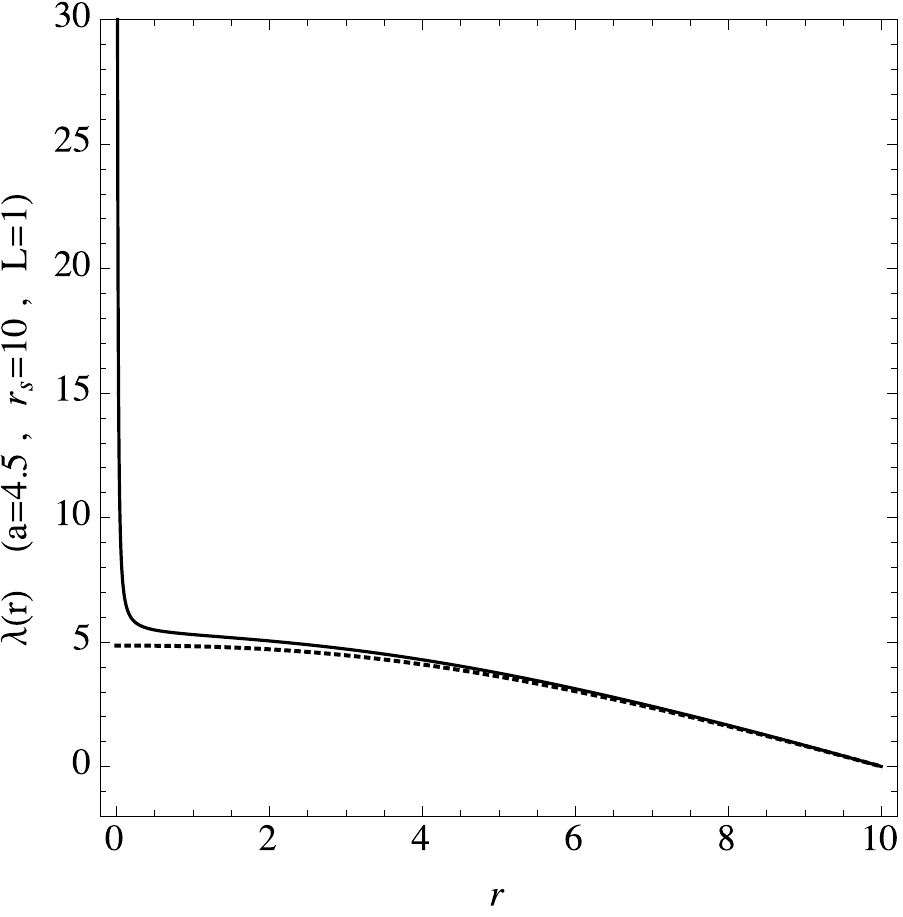}
\end{center}
\caption{Panel on the left: Plot of the affine parameter $\lambda(r)$ for an infalling radial motion of a massless particle. The affine parameter $\lambda \rightarrow + \infty$ for $r\rightarrow 0$, and therefore the rescaled Kerr spacetime for $r>0$ is geodesically complete. The dashed line shows the situation for not rescaled Kerr metric. Here we employ the following values for the parameters and the conserved quantities: $L=1$, $r_0=10$, $a=0.8$, $\theta=\pi/2$ and $r_s=10$. 
Panel on the right. It is again the affine parameter $\lambda(r)$, but for $a=4.5$.
\label{GeoPhotonsPlot}}
\end{figure}

In this last subsection about the geodesic completion of the conformally rescaled Kerr spacetime, we deal with massless particles. We repeat the analysis already applied to photons in the Schwarzschild metric discussed in~\ref{luceSch} to photons in the Kerr metric~(\ref{SKerr}). Once again, the stationarity (the metric is independent of $t$) and axial symmetry (the metric is independent of $\varphi$) imply the existence of the following two Killing vectors:
\be
\xi^{\alpha} = (1, 0, 0, 0) \, , \quad \eta^{\alpha} = (0, 0, 0, 1) \, .
\ee
Therefore, we have the following conserved quantities
\be
e = - \xi^\alpha u^\beta \hat{g}_{\alpha \beta} = - \left( \hat{g}_{tt} u^t + \hat{g}_{t \varphi} u^{\varphi} \right)\, , \quad 
\ell =  \eta^\alpha u^\beta \hat{g}_{\alpha \beta} = \hat{g}_{\varphi t} u^t + \hat{g}_{\varphi \varphi} u^{\varphi} .
\label{conservaKerr}
\ee
For photons
\be
\hat{g}_{\alpha \beta} u^\alpha u^\beta = 0 \quad \Longrightarrow \quad
 \hat{g}_{tt} \dot{t}^2 + 2 \hat{g}_{t \varphi} \dot{t} \dot{\varphi} + \hat{g}_{rr} \dot{r}^2 + \hat{g}_{\varphi \varphi} \dot{\varphi}^2 = 0 \, , 
 \label{dsl}
\ee
where here the dot $\dot{}$ stands for the derivative with respect to the affine parameter $\lambda$. Solving the two equations in~(\ref{conservaKerr}) for $\dot{t}$ and $\dot{\varphi}$, we find 
\be
\dot{t} = \frac{e \, \hat{g}_{\varphi \varphi}  + \hat{g}_{\varphi t} \, \ell }{ \hat{g}_{\varphi t}^2 - \hat{g}_{tt}  \hat{g}_{\varphi \varphi } }
\, , \quad 
\dot{\varphi} = - \frac{e \, \hat{g}_{\varphi t}  + \hat{g}_{t t} \, \ell }{ \hat{ g}_{\varphi t}^2 - \hat{ g}_{tt}  \hat{g}_{\varphi \varphi } } \, .
\label{LtLphi} 
\ee
Replacing (\ref{LtLphi}) in (\ref{dsl}) we end up with the radial geodesic equation 
\be
\hat{g}_{rr} \dot{r}^2 + \frac{e^2 \hat{g}_{\varphi \varphi} + 2 e \hat{g}_{\varphi t} \ell +\hat{g}_{tt}
   \ell^2}{\hat{g}_{tt} \hat{g}_{\varphi \varphi} - \hat{g}_{\varphi t}^2} =0 \, .
\ee
Since we are interested in the radial motion, we assume no orbital angular momentum, and then the geodesic equation simplifies to 
\be
\hat{g}_{rr} \dot{r}^2 + \frac{e^2 \hat{g}_{\varphi \varphi} }{\hat{g}_{tt} \hat{g}_{\varphi \varphi} - \hat{g}_{\varphi t}^2} =0 \quad \Longrightarrow \quad 
\frac{r^3 S(r)^2}{r^3+r a^2 + r_s a^2}  \left( \frac{d r}{d \lambda} \right)^2  = e^2 \, .
\label{GeoPhotonsKerr}
\ee
Note that the function in front of $\dot{r}^2$ is strictly positive for $r>0$. We can integrate (\ref{GeoPhotonsKerr}) for a photon directed towards $r=0$, and the result for the affine parameter $\lambda$ reads
\be
\lambda = - e^2 \int_{r_0}^r \sqrt{\frac{r^3 S(r)^2}{r^3+r a^2 + r_s a^2}} dr .
\ee
From the plot in Fig.~\ref{GeoPhotonsPlot}, it is clear that $\lambda \to + \infty$.

\section{The Raychaudhuri equation in conformal gravity} 
 
The Raychaudhuri equation for the expansion parameter has a purely kinematical meaning. Given the spacetime metric and the geodesic equation for massive probes, conformally coupled probes, or  photons, the Raychaudhuri equation has exactly the same form. Moreover, we know the spacetime metric and geodesic equations on the rescaled spacetime. Hence, we can directly evaluate the expansion parameter $\Theta$, which is the key quantity in the Raychaudhuri equation.

\subsection{Expansion parameter for massive particles conformally and non-conformally coupled in the non-singular Schwarzschild metric} 

The expansion parameter is defined as
\be
\Theta = u^\alpha;_\alpha = \frac{1}{\sqrt{|\hat{g}|} } \left(  \sqrt{|\hat{g}|} u^\alpha \right)_{,\alpha} \, .
\label{generalTheta}
\ee
We remind that for a conformally coupled particle falling into the black hole (\ref{Es}) 
\be
u^r \equiv \frac{ dr}{d \tau} = - \sqrt{ \frac{2 M}{r S(r)}} \, . 
\ee
Therefore, the $\Theta$ parameter for the congruence of geodesics of massive conformally coupled particles falling into  the non-singular spherically symmetric black hole reads
\be
\Theta_{\rm cp}  = -\frac{3 \left(r^2-3 L^2\right) \sqrt{2 M r}}{2 \left(L^2+r^2\right)^2} \, .
\ee
The above result shows that there is no focusing of the radial geodesics for $r \rightarrow 0$ as opposite to the Schwarzschild metric. When the limit $L\rightarrow0$ is taken, the singularity in $r\to0$ is recovered and geodesics are infinitely focused ($\Theta\to-\infty$):
\be
\lim_{L\rightarrow 0} \Theta_{\rm cp} = - \frac{3}{2} \sqrt{\frac{2 M }{r^3}} = \Theta_{\rm Schw} \, . 
\label{thetaSch}
\ee

We can also evaluate the expansion parameter $\Theta_{m}$ for a congruence of geodesics of massive  non-conformally coupled particles. We have just to replace in (\ref{generalTheta}) the radial geodesic equation (\ref{Es}) by the one   for a  non-conformally coupled massive particle (\ref{rdot2M}). The final result is
\be\label{thetamm}
\Theta_{m}= \frac{L^2 r^4 (2 {e} r+3 M)+r^6 (-2 ({e}-1) r-3 M)+L^6 (9 M-4 r)+3 L^4 r^2
   (5 M-2 r)}{\left(L^2+r^2\right)^3 \sqrt{\frac{ e r^5+\left(L^2+r^2\right)^2
   (2 M-r)}{r}}} \, .
\ee
If $L\neq0$, (\ref{thetamm}) does not develop singularity at $r=0$ and reduces again to (\ref{thetaSch}) in the limit $L\rightarrow 0$.

\subsection{Expansion parameter for photons in non-singular Schwarzschild metric}

The expansion parameter for photons $\Theta_l$ represents the fractional rate of change of the congruence's cross-sectional area $A$, namely 
\be
\Theta_l = \pm \frac{1}{\delta A} \frac{d \delta A}{d \lambda} \, ,
\label{AA}
\ee
where $\delta A$ is measured in the transverse directions. We consider the family of radially infalling light rays. For the Schwarzschild metric $\Theta_{l,{\rm Schw}} \rightarrow - \infty$ for $r\rightarrow 0$, while for the non-singular rescaled metric $\Theta_l$ tends to zero. Therefore, we have an infinite ``cross section" for photons on the Schwarzschild spacetime, but a vanishing cross section for photons on the non-singular black hole metric.  Roughly speaking, the infinite transverse area makes impossible for photons to scatter when approaching the point $r=0$.

Let us now explicitly evaluate the expansion parameter (\ref{AA}) for photons. We know that for photons 
\be
S^2(r) \left(\frac{d r}{d \lambda}\right)^2 = e^2 \, ,
\ee
and the area of the two-sphere is $4 \pi r^2 S(r)$, therefore the expansion parameter for ingoing photons reads \cite{Poissonbook}:
 \be
\Theta_l =   \frac{1}{\delta A} \frac{d \delta A}{d \lambda}  =  \frac{1}{4 \pi S(r) r^2}  \left(- \frac{d r}{d \lambda}\right)
\frac{d}{d r} \left(  4 \pi r^2 S(r)    \right) = - \frac{1}{4 \pi S(r) r^2} \frac{e}{S(r)} \frac{d}{d r} \left(  4 \pi r^2 S(r)    \right) = - \frac{2 e  r^3 \left(r^2-L^2\right)}{\left(L^2+r^2\right)^3} \, ,
\label{AA2}
\ee
which goes to zero for $r\rightarrow 0$. Since photons never reach $r=0$, the radial null geodesics 
become parallel when the affine parameter becomes infinite. The result of this computation agrees with the standard definition of the expansion parameters for null geodesics, that is $\Theta_l=k^\alpha_{;\alpha}$, where $k^\alpha$ is the photon four-velocity field.


\section{Conclusions and Remarks}

In this paper, we have explicitly showed that a large class of spacetime singularities are just an artifact of the conformal gauge in a general Weyl-invariant (or conformally invariant) gravitational theory. Singular and regular spacetimes are peculiar points on the same gauge orbit. Therefore, by a conformal rescaling of the metric, we can always move from one to another point of the same conformal orbit and the new metric turns out to be singularity-free. In short, in a Weyl conformally invariant gravitational theory 
characterized 
by the metric and the dilaton field we use such symmetry 
to move the singularity of a the spacetime into the unobservable dilaton field by means of a conformal factor.

We have studied the Schwarzschild and Kerr metrics, but likely our conclusions could be generalized to any singular metric, which is a solution of Einstein's gravity and the resolution of its singularity can be obtained in any conformally invariant theory. Other spacetimes, like FRW (which is trivially singularity-free in a conformally invariant theory) and the Kasner spacetime, have been analyzed in a previous work~\cite{CGLL}.

Our claims are based on the regularity of any curvature invariant and on the geodesic completeness of the spacetime. Indeed, we explicitly prove that massive particles (conformally or non-conformally coupled to gravity) and massless particles can never reach the former Schwarzschild singularity or the former Kerr ring singularity in a finite amount of proper time or other affine parameter, which characterizes their geodesics. All this leads us to claim that eventually a black hole looks like a ``bottomless black pit''.

Finally, there is a remarkable similarity between our regular black hole metrics and other singularity-free spacetimes discussed in the literature \cite{Bronnikov:2006fu, Bronnikov:2005gm}. The mathematical, but also physical, reason for the resolution of the Schwarzschild black hole spacetime singularity lies in the minimal area of the two dimensional sphere. A similar behavior, though somehow  more involved, is observed in the axially symmetric case. This resembles the geometry of wormholes and regular black holes, such as ``black universes'', discussed in \cite{Bronnikov:2005gm, Bronnikov:2006fu} (in the latter paper, it is provided a classification of spherically symmetric regular black holes). In particular, in~\cite{Bronnikov:2005gm} the authors started with the action for a self-gravitating minimally coupled scalar field with an arbitrary potential and for the case of a phantom field they derived a transparent analytic example of black hole metric with a minimal two-dimensional area.


\begin{acknowledgments}
C.B. acknowledges support from the NSFC (grants 11305038 and U1531117), Fudan University (Grant No.~IDH1512060), the Thousand Young Talents Program, and the Alexander von Humboldt Foundation.
\end{acknowledgments}


\appendix

\section{Kretschmann invariant for the non-singular Kerr metric \label{ap-1}}

The expression of the Kretschmann invariant for the singularity-free rescaled Kerr metric with the conformal factor $S(r)$ given in \eqref{grazieModeK} is
\be
&& \hspace{-0.3cm}
\hat K= \frac{}{\left(L^2+r^2+a^2 x^2\right)^{12}} \times 
\left(-3 a^{14} M^2 x^{14}+a^{12} \left(4 \left(23 x^4-60
   x^2+44\right) L^4-12 M^2 x^4 L^2+33 M^2 r^2 x^4\right) x^8
   \right. \nonumber \\
  && \hspace{-0.3cm}
  \left.
   +a^{10} \left(-8 \left(7 x^4-6
   x^2-8\right) L^6+2 \left(-9 M^2 x^4-8 M r x^2+4 r^2 \left(7 x^4-90 x^2+88\right)\right)
   L^4+144 M^2 r^2 x^4 L^2+117 M^2 r^4 x^4\right) x^6
   \right.  \nonumber  \\ 
   &&\hspace{-0.3cm}
    \left. 
   +a^8 \left(4 \left(23 x^4-48
   x^2+32\right) L^8-4 \left(3 M^2 x^4+8 M r \left(6 x^2-5\right) x^2+6 r^2 \left(5 x^4-4
   x^2-8\right)\right) L^6
   \right. \right.  \nonumber  \\
   && \hspace{-0.3cm}
   \left. \left. 
   +2 r^2 \left(189 M^2 x^4+8 M r \left(57 x^2-40\right) x^2+r^2
   \left(-302 x^4-240 x^2+528\right)\right) L^4+324 M^2 r^4 x^4 L^2+81 M^2 r^6 x^4\right)
   x^4
   \right.  \nonumber  \\
 &&\hspace{-0.3cm}
   \left. 
   +a^6 \left(\left(-3 M^2 x^4+16 M r \left(11-12 x^2\right) x^2+64 r^2 \left(x^4-3
   x^2+4\right)\right) L^8+8 r^2 \left(57 M^2 x^4+24 M r x^2+r^2 \left(24-10
   x^4\right)\right) L^6
   \right. \right.   \nonumber  \\
   && \hspace{-0.3cm}
   \left. \left. 
   +4 r^4 \left(15 M^2 x^4+24 M r \left(24 x^2-19\right) x^2-4 r^2
   \left(71 x^4-30 x^2-44\right)\right) L^4-81 M^2 r^8 x^4\right) x^2
   \right.   \nonumber \\
   && \hspace{-0.3cm}
   \left. 
   +r^6 \left(\left(531
   M^2-432 r M+92 r^2\right) L^8+4 r^2 \left(27 M^2+24 r M-14 r^2\right) L^6+2 r^4 \left(273
   M^2-216 r M+46 r^2\right) L^4 
   \right. \right.  \nonumber \\
   &&  \hspace{-0.3cm}
   \left. \left. 
   +12 M^2 r^6 L^2+3 M^2 r^8\right)-a^2 r^4 \left(\left(-64
   \left(x^2+3\right) r^2+496 M r+333 M^2 x^2\right) L^8+8 r^2 \left(3 \left(5 x^2+2\right)
   r^2-8 M \left(9 x^2-2\right) r
   \right. \right. \right.  \nonumber  \\
   &&\hspace{-0.3cm}
    \left. \left. \left. 
   +105 M^2 x^2\right) L^6-2 r^4 \left(4 \left(7 x^2+30\right)
   r^2-8 M \left(24 x^2+37\right) r+171 M^2 x^2\right) L^4+144 M^2 r^6 x^2 L^2+33 M^2 r^8
   x^2\right)
   \right.  \nonumber  \\
   && \hspace{-0.3cm}
   \left. 
   -a^4 r^2 \left(\left(-189 M^2 x^4-80 M r \left(3 x^2-4\right) x^2+8 r^2 \left(7
   x^4-24 x^2-16\right)\right) L^8+16 r^2 \left(30 M^2 x^4+6 M r \left(1-7 x^2\right) x^2
    \right. \right. \right.  \nonumber \\
   && \hspace{-0.3cm}
   \left. \left. \left. 
   +r^2
   \left(5 x^4+6 x^2-4\right)\right) L^6+4 r^4 \left(135 M^2 x^4-8 M r \left(45 x^2-56\right)
   x^2+r^2 \left(151 x^4-180 x^2-44\right)\right) L^4+324 M^2 r^6 x^4 L^2
   \right. \right.   \nonumber  \\
   && \hspace{-0.3cm}
   \left. \left. 
   +117 M^2 r^8
   x^4\right)\right) \, ,
\ee
where $x = \cos \theta$. 

\section{Radial geodesic equation for general values of the orbital angular momentum in axi-symmetric spacetime \label{ap-2}}

The radial geodesic equation in the equatorial plane ($\theta = \pi/2$ and $\dot{\theta}=0)$ for a massive non-conformally coupled particle for general values of angular momentum $\ell$ is:
\be
 \frac{1}{2} \dot{r}^2 
+ \underbrace{ \frac{1}{2} \frac{1}{\hat{g}_{rr}} \left[ \frac{e^2 \hat{g}_{\varphi \varphi} + 2 e \ell \hat{g}_{\varphi t} + \hat{g}_{tt} \ell^2}{\hat{g}_{t t} \hat{g}_{\varphi \varphi} - \hat{g}_{\varphi t} } + 1 \right]  + \frac{e^2 -1}{2}}_{V_{\rm eff}} = \frac{e^2-1}{2}  \, .
\label{NKell}
\ee
When the non-singular rescaled Kerr metric is inserted in~(\ref{NKell}), we end up with the following equation
\be
 && \hspace{-1cm} 
 \frac{1}{2} \dot{r}^2 
+ V_{\rm eff} = \frac{e^2-1}{2} \,\quad\quad \text{for the effective potential given by}\\
&& \hspace{-1cm}
V_{\rm eff} = \frac{a^2 \left( -  \left(e^2 (r +  r_s )-r S(r) \right)\right)+2 a e \ell  {r_s}+r^2
   \left(e^2 r \left( S(r)^2 - 1\right)+S(r) (-r (S(r)-1)- {r_s})\right)+ \ell^2 (r-{r_s})}{2
   r^3 S(r)^2} \, .
   \ee


\end{document}